\documentclass[conference, two column]{IEEEtran}
\usepackage{multirow}
\usepackage{array}
\usepackage{cite}
\usepackage{amsmath,amssymb,amsfonts}
 
\usepackage{algorithmic}
\usepackage{graphicx}
\usepackage{textcomp}
\usepackage{comment}
\usepackage{color}
\usepackage{epstopdf}
\usepackage{listings}
\usepackage{amsthm}

\usepackage{float}
\usepackage{xcolor}
\usepackage[center]{caption} 
\usepackage{subcaption}
\usepackage{color} 
\definecolor{mylilas}{RGB}{170,55,241}
\usepackage{float} 

\usepackage{caption} 
\usepackage{multicol}
\usepackage{mathtools} 
\usepackage{cuted} 
\usepackage[ruled,vlined]{algorithm2e}
\usepackage{hyperref}

\usepackage{lipsum}
\usepackage{enumitem}
\usepackage[switch]{lineno}

\begin{document}\title{Bearing-only Tracking using Towed Sensor-Array with Non-Gaussian Measurement Noise Statistics  }
\author{
\IEEEauthorblockN{Rohit Kumar Singh\IEEEauthorrefmark{1}\IEEEauthorrefmark{3}, Subrata Kumar \IEEEauthorrefmark{2}, Shovan Bhaumik \IEEEauthorrefmark{3}}\\
\IEEEauthorblockA{\IEEEauthorrefmark{1} Department of Electronics and Communication Engineering, Pandit Deendayal Energy University, Gandhinagar, India}
\IEEEauthorblockA{\IEEEauthorrefmark{2}Department of Mechanical Engineering, Indian Institute of Technology Patna, India}
\IEEEauthorblockA{\IEEEauthorrefmark{3}Department of Electrical Engineering, Indian Institute of Technology Patna, India}

\IEEEauthorblockA{\IEEEauthorrefmark{1} \emph{rohit\_1921ee19@iitp.ac.in,}\IEEEauthorrefmark{2} \emph{subrata@iitp.ac.in,}\IEEEauthorrefmark{3} \emph{shovan.bhaumik@iitp.ac.in}}

}
	\maketitle
    \begin{abstract}
Passive bearing-only tracking (BOT) estimates the target states by utilising noisy bearing measurements captured by a sensor array.
The sensor array is often towed behind the ship, using a long flexible cable to reduce interference from the own-ship's inherent noises. This forms a towed cable sensor-array system (TCSAS). 
During BOT, the tow-ship has to perform a manoeuvre to make the tracking system observable.
Such a manoeuvre destabilises the TCSAS, thus making its exact location unknown \emph{w.r.t.} tow-ship.
However, it is very crucial to know the exact location of the towed sensor-array to perform efficient and reliable target state estimation.
The existing BOT approaches perform TMA during own-ship manoeuvre either by pausing the measurement updation step of the estimation algorithm or assuming a fixed aft position for the towed sensor-array. These assumptions lead to unreliable state estimation.
To address this, we propose a dynamic model for TCSAS, using a lumped mass approach, which will provide the location of the sensor array during the own-ship manoeuvre. This location will be fed to the state estimation algorithm. The dynamic of TCSAS in 3D space is obtained by solving the equations obtained from the moment balance condition and quasi-static equilibrium condition at the lumped mass points. 
Moreover, the bearing data captured by the towed sensor-array is corrupted with non-Gaussian noise. It is handled using the maximum correntropy criterion based Kalman filter with a kernel bandwidth selection technique, proposed in this paper.
The proposed sensor-array dynamic model is verified for a real-world BOT engagement scenario.

    \end{abstract}

	\section{Introduction}

Target tracking is a process of estimating the states of a target using noisy measurements such as bearing angle \cite{das2024passive}, Doppler frequency \cite{ding2020human}, and time delay of arrival \cite{alexandri2021time}. This work focuses on passive bearing-only tracking (BOT), where a single own-ship, moving at a constant velocity, tracks a target using the bearing data collected by the hydrophone sensor \cite{singh2022passive}.

The relationship between the measured bearing data and state vectors of own-ship and target is nonlinear, which makes it a nonlinear state estimation problem. The state estimation is performed in the Bayesian filtering framework, where the Chapman-Kolmogorov equation predicts the prior state estimate, and Bayes' rule updates on arrival of measurement data to obtain the posterior estimate \cite{bhaumik2019nonlinear}. Solving these equations for nonlinear estimation problem requires evaluating intractable integrals \cite{kumar2023new}.
As the integrals are intractable, they have been approximately evaluated in extended Kalman filter (EKF) using a first-order Taylor series expansion \cite{bar2004estimation} or by utilising a few support points and associated weights in deterministic sample point filters (DSPF). Based on the support point generation methods, different filters such as unscented Kalman filter (UKF) \cite{van2001square}, cubature Kalman filter (CKF) \cite{arasaratnam2009cubature}, Gauss-Hermite filter (GHF) \cite{ito2000gaussian}, polynomial chaos Kalman filter (PCKF) \cite{kumar2023polynomial}, and many more are proposed.
These filters are based on assumptions that the prior and posterior pdfs of state follow Gaussian distribution \cite{kumar2023new}.

\begin{figure}
	\centerline{\includegraphics[scale =0.5]{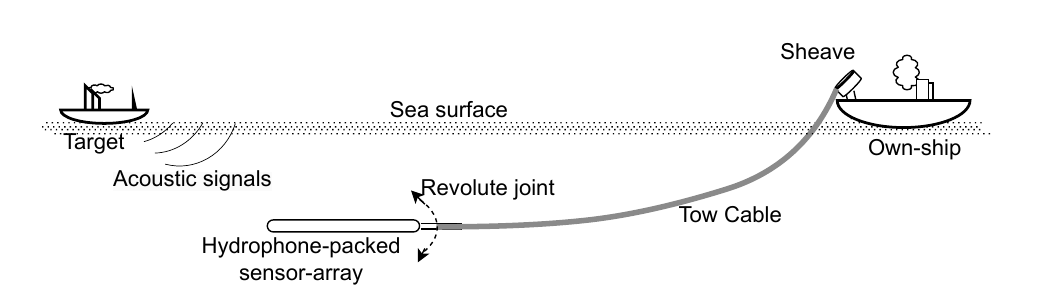}}
	\caption{Bearing-only tracking with bearing data collected using towed cable sensor-array system.}
	\label{figBOT}
\end{figure}

In passive target motion analysis (TMA) using BOT method, a hydrophone sensor passively listens to incoming acoustic signals from the target, ensuring that the ship’s location remains undisclosed, providing an advantage in covert operations \cite{radhakrishnan2018gaussian}. This sensor can be either hull-mounted or towed behind the ship using a long cable \cite{lemon2004towed}. The problem with the hull-mounted sensor-array is that the inherent ship noise is too loud, which interferes and distorts the faint acoustic signal of interest coming from target to own-ship \cite{abraham2019underwater}. To tackle this issue, the sensor-array is towed behind the tow-ship using a long cable as shown in Fig. \ref{figBOT}. This creates a separation of the sensor-array from own-ship's inherent noise, such that the sensor-array can easily capture the faintest of signals incoming from the target lying within it's detection range \cite{lemon2004towed,yang2023dynamic}.

The bearing data is collected from a single tow-ship, a process commonly known as an autonomous TMA. For such a process, the target moving with a constant velocity will make the system unobservable \cite{song1996observability}. To ensure observability, the tow-ship must perform a manoeuvre \cite{radhakrishnan2018gaussian,leong2013gaussian}. 
This tow-ship manoeuvre destabilizes the towed sensor array, making its exact position untraceable. 
A major challenge in performing BOT with towed cable sensor-array system (TCSAS) is that the state estimation algorithm for effective TMA requires precise knowledge of the sensor-array's location relative to the tow-ship \cite{zhang2023dynamic}. So, it becomes essential to understand the dynamics of the TCSAS to accurately determine the sensor array’s position during own-ship manoeuvres. The existing literature on BOT using TCSAS tackles the uncertainty in sensor-array location during the own-ship manoeuvre by either halting the measurement updation of the estimation process during the manoeuvre \cite{la2008analysis} or by performing the estimation using the approximated location of towed sensor-array \cite{kumar2021conditioned, jahan2020fusion}. Previous works on BOT using towed sensor-array presented in \cite{kumar2021conditioned,jahan2020fusion,kumar2016integrated,xu2017particle,kim2022surface,lebon2021tma} assumed that the TCSAS maintains precise linear alignment with the vessel’s heading, remaining directly aft during own-ship manoeuvre. These existing approaches didn't consider the dynamics of the towed array during TMA, leading to unreliable estimations.
Additionally, integration of positioning sensors directly onto towed sensor arrays for its localization is also explored in literature \cite{zhang2023autonomous, lemon2004towed}. However, the lack of GPS signals in underwater environments makes this approach particularly challenging.

So, during the tracking interval, an accurate location of the sensor-array is required without employing any additional sensor. Unfortunately, there is no existing literature available that models the TCSAS. But, the literature on dynamic models of towed marine cable \cite{sanders1982three,huang1994dynamic,guo2021numerical} with attached tug-boat \cite{feng2022study,sun2011dynamic} and remotely operated vehicle \cite{driscoll2000development} are available. They are based on a lumped mass approach, where the long cable is discretized into a finite number of interconnected segments, with the mass of each segment considered to be concentrated at the lumped mass point \cite{calnan2018reference, driscoll2000development, sanders1982three, feng2022study, guo2021numerical}. The force balance condition at the lumped mass nodes yields a non-linear ordinary differential equation (ODE) which is solved by finite difference method to generate the location and orientation of lumped mass point and towed-cable segments \cite{huang1994dynamic, sun2011dynamic, zhu2023efficient, sanders1982three}.
These existing dynamic models for towed marine cables use force balance conditions at the interconnected nodes. But, these approaches do not account for the incorporation of the rotational inertia effects of tow-ship motion on the towed body \cite{liu2023study,zhu2023efficient,sanders1982three,huang1994dynamic,zhang2023dynamic}. 
In our earlier work \cite{singh2025dynamics}, a dynamic model for the TCSAS using three interconnected rigid segments was developed. In \cite{singh2024bearing}, the bearing data obtained from this model were utilised to perform BOT.

In this paper, we propose a dynamic model of TCSAS based on Newtonian mechanics, where the azimuth orientation and position of the lumped mass segments in the $xy$ plane are determined using moment balance conditions at the interconnected lumped mass nodes. This ensures that the dynamic model incorporates rotational inertia effects of tow-ship’s motion on the towed body. Additionally, elevation angle and depth of the interconnected segments are obtained from quasi-static equilibrium conditions at lumped mass points in the $z$-plane. The proposed approach provides $3D$ modelling of the TCSAS, where the cable is modelled as flexible using a user-defined number of interconnected segments.
The cable is hinged to the stern of tow-ship, and sensor-array is attached to the end of the cable using a revolute joint as shown in Fig. \ref{figBOT}.
The sensor-array maintains a horizontal alignment with water surface using revolute joint, which is essential for effective beamforming \cite{stiles2013dynamic}. 
The proposed dynamic modelling provides an accurate location of the towed sensor-array for any type of tow-ship manoeuvre.
 The obtained position is fed into the state estimation algorithm, enabling reliable and accurate target state estimation during BOT.




The second challenge encountered in target state estimation using BOT is that the bearing data collected by sensor-array is corrupted with noise having non-Gaussian pdf \cite{chen2017maximum,liu2018maximum}. Existing literature for passive BOT primarily assumes Gaussian noise in sensor-array data and their states are estimated using Gaussian filters\cite{kumar2023new,kumar2023polynomial, kumar2021conditioned, jahan2020fusion,kumar2016integrated, xu2017particle, kim2022surface}. However, real-world situations often involve impulsive disturbances and non-Gaussian noise. The Gaussian filters mentioned earlier, such as the EKF, UKF, CKF, GHF, and PCKF, rely on the minimum mean square error (MMSE) criterion. Such that these filters effectively capture up to second-order moments of noise distribution. However, in presence of impulsive and non-Gaussian noises, MMSE-based filters fail to account for higher-order moments, leading to degraded estimation performance \cite{singh2010closed}.
The existing literature tackles the non-Gaussian measurement noise utilizing an information-theoretic learning quantity called correntropy \cite{chen2017maximum}. It can capture both second-order and higher-order even moments of the estimation error \cite{liu2007correntropy}, thus effectively handling the non-Gaussian noise and improving target state estimation accuracy \cite{liu2007correntropy,liu2016extended,liu2018maximum,hou2017maximum}.

The kernel function plays a crucial role in maximum correntropy (MC)-based filtering methods. The Gaussian kernel is widely used due to its smoothness, symmetry, and the property that the product of two Gaussian functions results in another Gaussian function, which simplifies analytical computations \cite{liu2007correntropy}. While alternative kernels like Laplacian \cite{hu2021robust} and Gaussian mixture kernels \cite{wang2022mixture} have been explored, the Gaussian kernel remains preferred for its mathematical tractability.
In MC filters, the kernel bandwidth is a crucial parameter that determines the sensitivity of the filter to measurement discrepancies, making its appropriate selection crucial for achieving accurate state estimation \cite{saha2023robust}. While some studies have explored adaptive kernel bandwidth selection methods in \cite{cinar2012hidden, hou2017maximum, fakoorian2019maximum}. These approaches often lack guarantees for optimality and are designed for specific problems. 
In this work we propose a kernel bandwidth selection technique, which involves evaluating the posterior error covariance for a range of bandwidth values and selecting the one that minimizes the trace of the posterior error covariance matrix.
       

In summary, this study addresses two key challenges in target state estimation during passive BOT. The first challenge is accurately determining the location of the towed sensor-array during the tow-ship manoeuvre. This is resolved through the proposed dynamic modeling of the TCSAS, which involves solving equations derived from the moment balance conditions and quasi-static equilibrium at the lumped mass points. The second challenge is handling non-Gaussian noise in sensor-array data. It is tackled using maximum correntropy criterion-based Kalman filtering in the Bayesian framework with a novel kernel bandwidth selection technique.
To validate the proposed approach, a real-world BOT engagement scenario is considered \cite{radhakrishnan2018gaussian,leong2013gaussian,das2024passive}, where bearing data is collected from the proposed dynamically modelled towed sensor-array. Target state estimation is then performed using the maximum correntropy-based filtering technique. The results are compared with existing BOT methods that assume a fixed aft position for the towed sensor-array \cite{kumar2016integrated,kumar2021conditioned,jahan2020fusion,lebon2021tma}, demonstrating the advantages of the proposed approach in improving estimation accuracy.
  
This paper is further organized as follows: Section II presents the problem formulation for BOT. Section III develops the mathematical model for the TCSAS using moment balance and quasi-static equilibrium conditions. Section IV describes a generalized maximum correntropy Kalman filtering method in the Bayesian framework with a novel kernel bandwidth selection technique, and Section V presents simulation results for a BOT engagement scenario, where target states are estimated using maximum correntropy-based filter utilizing the sensor-array data collected from the proposed dynamic model of TCSAS. 

\section{Problem Formulation}
The objective of this work is to track a sea-surface target moving at a nearly constant velocity using bearing data collected from a sensor-array towed behind a ship.
The position and velocity of the target at $k^{th}$ time instant are taken as $(x_{t,k},y_{t,k})$ and $(v_{t,x,k},v_{t,y,k})$, respectively and its state vector representation is $X_{t,k}=\begin{bmatrix}x_{t,k} & y_{t,k}  & v_{t,x,k} & v_{t,y,k}\end{bmatrix}^\top$. Similarly, the state vector representation of tow-ship is taken as $X_{o,k}=\begin{bmatrix}x_{o,k} & y_{o,k}  & v_{o,x,k} & v_{o,y,k}\end{bmatrix}^\top$ and centre of gravity of towed sensor-array is taken as $X_{s,k}=\begin{bmatrix}x_{s,k} & y_{s,k}  & v_{s,x,k} & v_{s,y,k}\end{bmatrix}^\top$, respectively. 
A relative state vector is defined as $X_{k} = X_{t,k}-X_{o,k}$. The target dynamics evolution can be represented using a state space difference equation given as \cite{radhakrishnan2018new}
\begin{equation} \label{eqnprocessmodel}
   X_{k} = FX_{k-1}-U_{k,k-1}+\mu_{k-1},
\end{equation}
where F is a state transition matrix given as $F=[ I_{2 \times 2} \ \; TI_{2 \times 2} ; \ \;  0_{2 \times 2} \ \; I_{2 \times 2}]$. $T$ is the sampling interval and $I$ is the identity matrix. $U_{k,k-1}=[
		x_{o,k}-x_{o,k-1}-Tv_{o,x,k-1} \ \; y_{o,k}-y_{o,k-1}-Tv_{o,y,k-1} \ \; v_{o,x,k}-v_{o,x,k-1}  \ \; v_{o,y,k}-v_{o,y,k-1}]^\top$ is the accelerating inputs to the tow-ship for directing its motion.
$\mu_{k-1}\sim \mathcal{N}(0,Q_{k-1})$ is independently identically distributed Gaussian noise with zero mean and $Q_{k-1}$ covariance, where $Q = \bar{q}\begin{bmatrix}
	\dfrac{T^3}{3}I_{2 \times 2} & \dfrac{T^2}{2}I_{2 \times 2} \ ;\  \dfrac{T^2}{2}I_{2 \times 2} & TI_{2 \times 2} 
\end{bmatrix}$. $\bar{q}$ is the process noise intensity.

The noise corrupted bearing measurements of the target \emph{w.r.t.} true north are obtained from the towed sensor-array. The bearing measurement at $k^{th}$ time instant is expressed as
\begin{equation} \label{eqnmeasurementmodel}
	Y_k = \Phi(X_{k})+r_{k},
\end{equation}
where $\Phi(X_{k})=\tan^{-1}(\dfrac{x_{t,k}-x_{s,k}}{y_{t,k}-y_{s,k}})$ is bearing angle. $r_k$ is non-Gaussian measurement noise, which is modelled as a weighted sum of Gaussian noise \cite{chen2017maximum,xu2021maximum}.

The effect of additive non-Gaussian noise in bearing data is addressed in state estimation algorithm using the maximum correntropy criterion. Additionally, the location of centre of gravity (CG) of towed sensor-array, which is assumed to be the point of bearing data acquisition, is determined through the dynamic modelling of TCSAS. 

\section{Mathematical model of towed cable sensor-array system}\label{sectiondynamicmodel}
\begin{figure}[ht]
	\centerline{\includegraphics[scale =0.5]{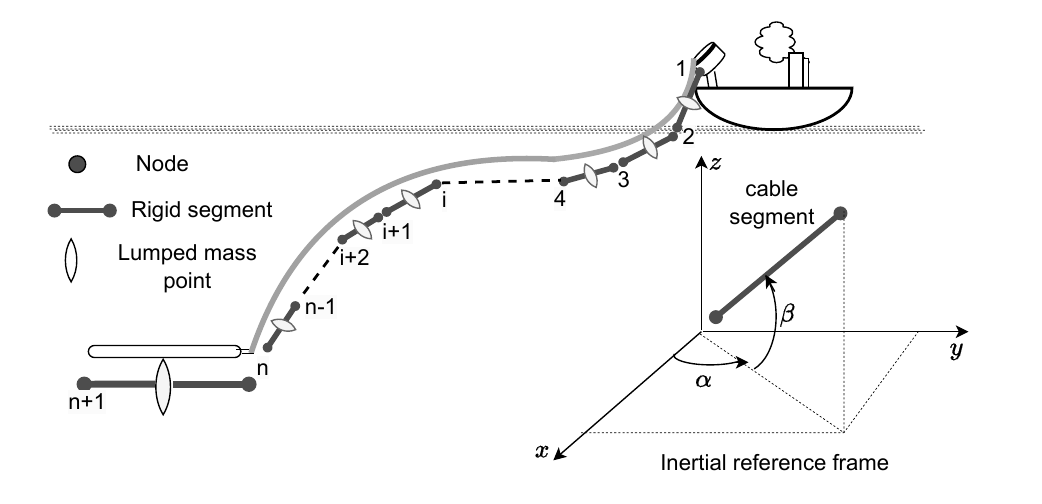}}
	\caption{Lumped mass model of TCSAS.}
	\label{figlumped}
\end{figure}
The towed system consists of tow-ship, flexible cable and a rigid sensor-array. The top end of the flexible cable is connected to the tow-ship using a sheave, and the sensor-array is connected at the bottom end of the cable \cite{lemon2004towed}. 
To perform the mathematical modelling of the TCSAS, the flexible cable is modelled using the lumped mass approach as shown in Fig. \ref{figlumped}. The cable is modelled flexible using (n-1) rigid segment rods, which are interconnected at n nodes \cite{liu2023study}. 
The roll motion of the towed body is not considered in this study so that the cable has 5 degree of freedom (DOF). 
The sensor-array is connected to the $n^{th}$ lumped mass node of cable using a revolute joint,
so that it maintains a horizontal alignment \emph{w.r.t.} horizontal plane which is essential for beam forming \cite{stiles2013dynamic}. Sensor-array has 4 DOF as its elevation angle is restricted at zero degree. 
The position of tow-ship is known for all the time instant. To deduce the position of towed cable segments and sensor-array \emph{w.r.t.} tow-ship position, requires the azimuth ($\alpha$) and elevation angle ($\beta$) information at each nodes for all time instant. 
The azimuth angle of the interconnected rigid segments changes when the tow-ship performs a manoeuvre, which is to be obtained by performing the dynamic analysis of towed system using the moment balance condition at the interconnected nodes \cite{guo2021numerical}. 
The elevation angle of each segment \emph{w.r.t.} horizontal $xy$ plane is obtained by solving the force balance equation deduced under quasi-static equilibrium condition at the interconnected nodes \cite{liu2023study,zhu2023efficient}, using this, the submersion depth of each rigid segment can be acquired.
Thus, the complete $3D$ modelling for TCSAS can be obtained using the moment balance condition and quasi-static equilibrium condition at the nodes of interconnecting segments for each time instant. 
 Before deriving the equations using moment balance and quasi-static equilibrium condition, there are some assumptions undertaken in this study for modelling the TCSAS, which are
\begin{itemize}
 \item The towed cable is cylindrical with uniform weight distribution.
 \item The effect of added mass on towed body during its transverse motion is not considered in the modelling. 
 \item The water is considered still of uniform density.
 \item The drag coefficient remains constant irrespective of angle of attack between the towed body and water.
 \item The axial elongation of the towed body is neglected.
\end{itemize}  
 \subsection{Moment balance condition at the nodes} \label{subsectionmoment}
  \begin{figure}[ht]
 	\centerline{\includegraphics[scale =0.48]{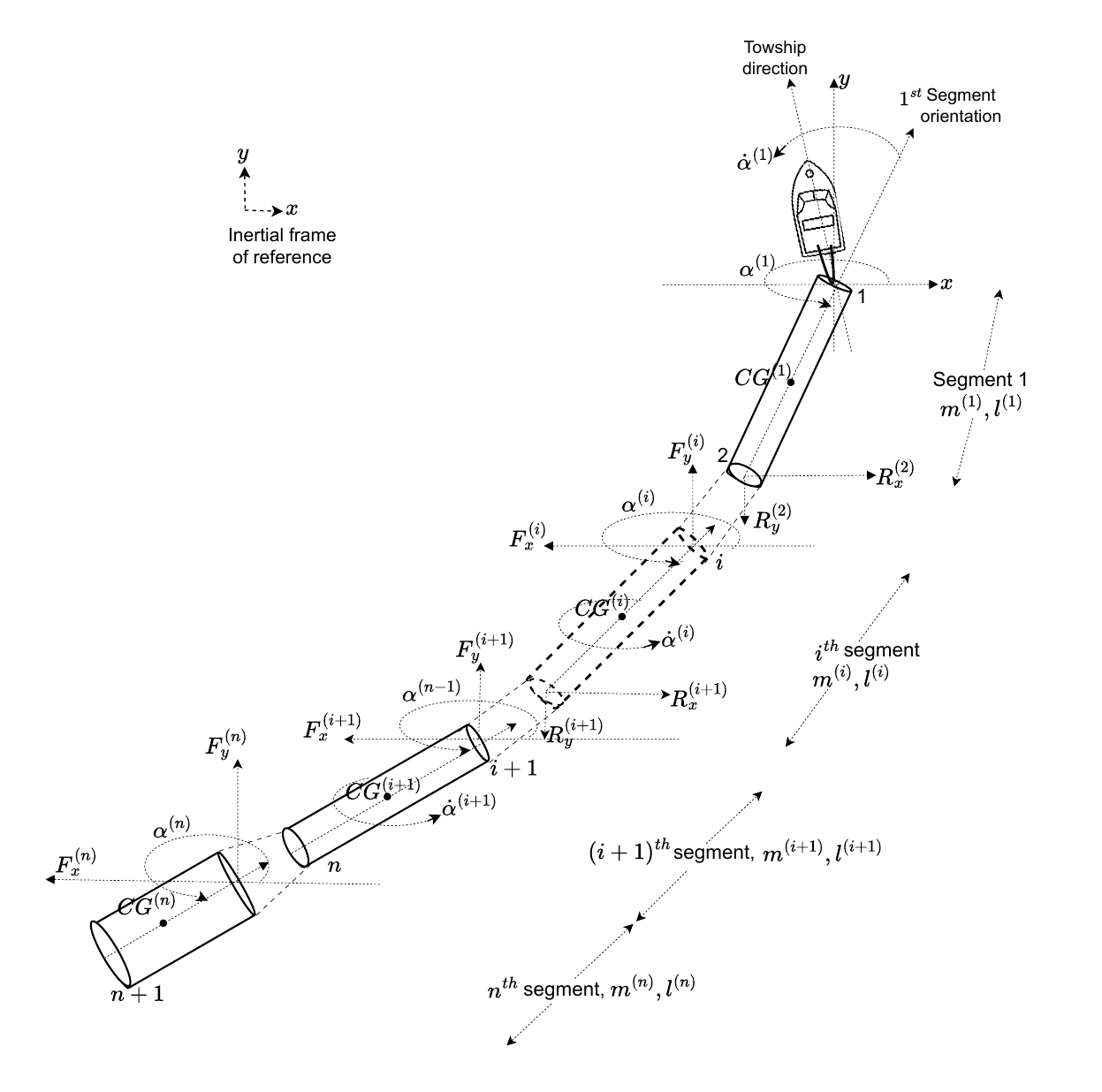}}
 	\caption{Free body diagram of TCSAS.}
 	\label{figfreebody}
 \end{figure} 
The tow-ship moves in $xy$ plane, and azimuth orientation of lumped mass modelled rigid segments changes during the tow-ship manoeuvre.
The azimuth orientation of interconnected rigid segments is obtained from the dynamic modelling analysis of projected TCSAS in $xy$ plane as shown in Fig. \ref{figfreebody}. 
It can be seen from Fig. \ref{figfreebody} that the cable is modelled using $(n-1)$ discretized rigid segments, and it is hinged to tow-ship at $1^{st}$ node. 
The position, velocity and acceleration at this first (top) node are identical to those of the tow-ship.
The sensor-array is connected to the cable end at $n^{th}$ node. The initial azimuth orientation for the $1^{st}$ $\cdots$ $i^{th}$, $i+1^{th}$, and $n^{th}$ rigid segment are $\alpha^{(1)}$ $\cdots$ $\alpha^{(i)}$, $\alpha^{(i+1)}$, and $\alpha^{(n)}$, respectively. Then the tow-ship performs a manoeuvre with an angular velocity of $\dot\alpha^{(1)}$, which subsequently changes the azimuths for each rigid segments as shown in Fig. \ref{figfreebody}. 
The updated azimuth orientation of the rigid segments is obtained by solving the equation deduced from the moment balance equilibrium condition at the interconnected nodes.
The dynamic modelling of projected TCSAS in $xy$ plane using the moment balance condition at the nodes of interconnected rigid segments incorporates the rotational inertia effect of tow-ship motion on the towed body.

The equation derived from the moment balance condition at $i^{th}$ node is given as \cite[pp. 417]{meriam2020engineering}
 \begin{equation} \label{eqmoment}
	\textstyle\sum M_{ext,k}^{(i)} =\dot{H}_{k}^{(i)} + \vec{r}_{CG}^{(i)} \times m^{(i)}\vec{a}_k^{(i)} +\textstyle\sum M_{damp,k}^{(i)},
\end{equation}
where $\textstyle\sum M_{ext,k}^{(i)}$ is the sum of moment generated due to external forces acting on $i^{th}$ segment at $k^{th}$ time instant. 
It is expressed as $\textstyle\sum M_{ext,k}^{(i)}=\textstyle\sum M_{D,k}^{(i)}-\textstyle\sum M_{reac,k}^{(i)}$.
$\textstyle\sum M_{D}^{(i)}$ is the moment due to hydrodynamic forces ($\mathcal{D}$) acting on $i$ rigid segment, and $\textstyle\sum M_{reac}^{(i)}$ is the moment due to reaction force ($R$) exerted by $(i+1)^{th}$ segment on $i^{th}$ segment at $i^{th}$ node. 
$\textstyle\sum M_{damp,k}^{(i)} = \mathcal{C}_{damp}\dot{\theta}_k^{(i)}$ is the rotational damping moment acting on node $i$ which models the cable internal deformation property which is used in stabilizing the rotational effect of tow-ship sharp manoeuvre on the interconnected rigid segments \cite{sun2011dynamic}, where $\mathcal{C}_{damp}=\dfrac{\pi\varrho d^3l^{(i)}}{3}$ is the rotational damping coefficient with $\varrho$, $d$ and $l^{(i)}$ as water viscosity, diameter and length of $i^{th}$ cable segment, respectively \cite[pp. 48]{rao1995mechanical}. $\dot H_{rel,k}^{(i)}=I^{(i)}\ddot\theta_k^{(i)}$ is the moment at $i^{th}$ node due to force acting at $i^{th}$ segment. $I^{(i)}=\dfrac{m^{(i)}l^{(i)2}}{3}$ is the moment of inertia for $i^{th}$ segment and $m^{(i)}$ as the mass of the $i^{th}$ segment. The term ($\vec{r}_{CG}^{(i)} \times m\vec{a}_k^{(i)}$) accounts for the moment generated at node $i$ due to inertial force from the tow-ship motion \cite{meriam2020engineering}. It is expressed as
\begin{equation}\label{eqninertial}
\begin{split}
		\vec{r}_{CG}^{(i)} \times m^{(i)}\vec{a}_k^{(i)} = & (\frac{l^{(i)}}{2} \cos\alpha_k^{(i)} \hat{i} + \frac{l^{(i)}}{2} \sin\alpha_k^{(i)} \hat{j}) \times (a_{x,k}^{(i)} \hat{i} \\ & +a_{y,k}^{(i)} \hat{j}),\\
		=  & \frac{m^{(i)}l^{(i)}}{2}(a_{y,k}^{(i)} \cos\alpha_k^{(i)} - a_{x,k}^{(i)} \sin\alpha_k^{(i)} ),
\end{split}\end{equation}
where ($a_{x}^{(i)},a_{y}^{(i)}$) are acceleration components of node $i$.
The moment balance equation at $i^{th}$ node for $k^{th}$ time instant, obtained by substituting Eqn. \eqref{eqninertial} into Eqn. \eqref{eqmoment} is given by
 \begin{equation}\label{eqmomentexpanded}\begin{split}
	\textstyle\sum M_{D,k}^{(i)}-\textstyle\sum M_{reac,k}^{(i)} = &\dfrac{m^{(i)} l^{(i)2}}{3}\ddot\alpha_k^{(i)} + \frac{m^{(i)} l^{(i)}}{2}(a_{y,k}^{(i)} \cos\alpha_k^{(i)} \\ & - a_{x,k}^{(i)} \sin\alpha_k^{(i)} ) +\dfrac{\pi\varrho d^{(i)3} l^{(i)}}{3}\dot{\alpha}_k^{(i)}.
\end{split}\end{equation}
 The drag force acting perpendicular to the body is referred to as pressure or normal drag, while the drag force acting tangential to the body is known as tangential or shear (friction) drag.  It is to be noted that the tangential drag force acts along the rigid segment, due to which it does not generate any moment \cite[ch. 9]{nakayama2018introduction}.
 As a result, the moment generated due to drag force at node $i$, \emph{i.e.} ($\textstyle\sum M_{D,k}^{(i)}$) is solely due to normal drag force, which is given as
 \begin{equation}\label{eqmomentdrag}
\textstyle\sum M_{D,k}^{(i)} = \int_{o}^{l^{(i)}} \vec{r}\times d\vec{\mathcal{D}}_{n,k} , 
\end{equation} 
 where $d\vec{\mathcal{D}}_{n}$ is the pressure (normal) drag force acting on fractional element $dr$ located at a distance of $r$ from the node $i$. 
 
 The moment due to reaction force exerted by $(i+1)^{th}$ segment on $i^{th}$ segment at $i^{th}$ node is given as
\begin{equation} \label{eqmomentreaction}
	\textstyle\sum M_{reac,k}^i =  R_{x,k}^{(i+1)}(l^i\sin{\alpha}_{k}^{(i)}) - R_{y,k}^{(i+1)}(l^{(i)}\cos{\alpha}_{k}^{(i)}), 
\end{equation}
where $R_{x}^{(i+1)}$ and $R_{y}^{(i+1)}$ are the reaction force components due to $(i+1)^{th}$ segment acting at $i^{th}$ node as shown in Fig. \ref{figfreebody}. These reaction forces are expressed as
\begin{equation} \label{eqreaction}\begin{split}
		R_{x,k}^{(i+1)} = 
         -(m^{(i)} a_{x,CG,k}^{(i+1)}
        -\mathcal{D}_{n,x,k}^{(i+1)} - \mathcal{D}_{t,x,k}^{(i+1)} + F_{x,k}^{(i+2)}), \\
		R_{y,k}^{(i+1)} = 
        -(m^{(i)} a_{y,CG,k}^{(i+1)} -\mathcal{D}_{n,y,k}^{(i+1)} - \mathcal{D}_{t,y,k}^{(i+1)} +  F_{y,k}^{(i+2)}),
\end{split} \end{equation}
where $\mathcal{D}_{n}^{(i+1)}$ and $\mathcal{D}_{t}^{(i+1)}$ are the normal and tangential drag force, respectively acting on $(i+1)^{th}$ segment.
It can be observed from Eqns. \eqref{eqmomentdrag} and \eqref{eqreaction} that the moment generated due to normal drag and reaction force necessitates an expression for the drag force acting on tow body. Hence, a generalised expression for the drag force components acting on a single rigid towed segment is derived below. This expression will be subsequently used to obtain the complete form of the moment balance Eqn. \eqref{eqmomentexpanded} at $i^{th}$ node.
  
\textbf{Drag force acting on a single towed body:}
 \begin{figure}[ht]
	\centerline{\includegraphics[scale =0.5]{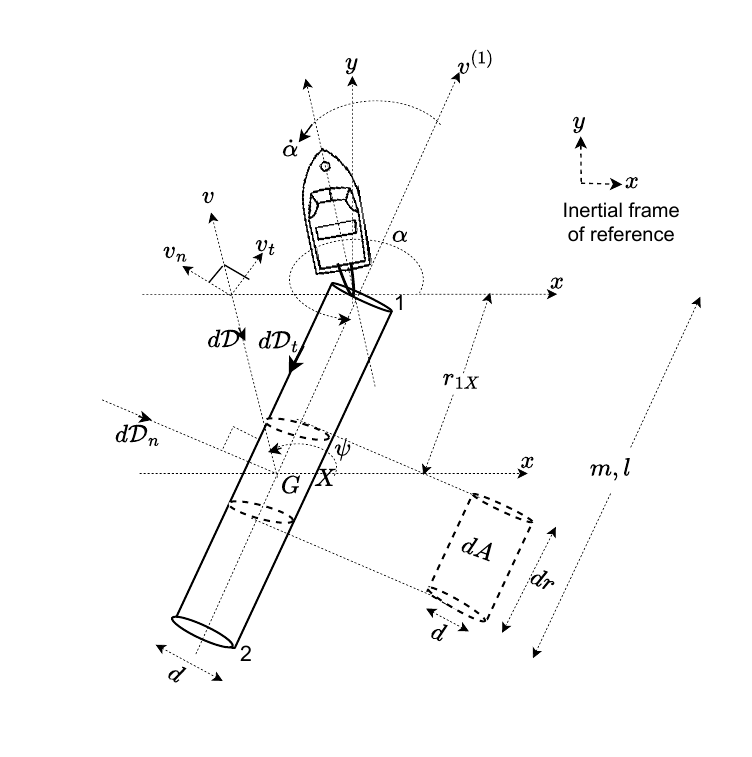}}
	\caption{Free body diagram of single towed segment in $xy$ plane.}
	\label{figdragmodel}
\end{figure} 
A rigid segment of mass $m$, length $l$ and diameter $d$ is towed to a surface ship moving with a velocity $v_1$ as shown in Fig. \ref{figdragmodel}. 
A fractional element X of rigid segment, having an interacting surface area $dA=d\times dr$ with the fluid is considered for drag calculation.
The tow-ship performs a turn with angular velocity $\dot\alpha$, and the velocity of fractional element $X$ is considered as $v$.
The drag force $d\mathcal{D}$ acting on fractional element is in opposite direction to velocity $v$ at an angle $\psi$ \emph{w.r.t.} x-axis as shown in Fig. \ref{figdragmodel}. The normal ($d\mathcal{D}_n$) and tangential ($d\mathcal{D}_t$) drag force acting on a fractional rigid segment is given by Morison equation \cite[ch.9]{nakayama2018introduction} as
\begin{equation}\label{eqdragfractional}\begin{split}
		d\mathcal{D}_{n} = & -\dfrac{1}{2} \rho C_{D,n} d \vec{v}_{n} |\vec{v}_{n}|dr,\\
		d\mathcal{D}_{t} = & -\dfrac{1}{2}\rho C_{D,t} \pi d \vec{v}_{t} |\vec{v}_{t}|dr,
\end{split}\end{equation} 
where $\rho$ is the density of fluid. $C_{D,n}$ and $C_{D,t}$ are the normal and tangential drag coefficient, respectively. $\vec{v}_n$ and $\vec{v}_t$ are the normal and tangential component of velocity $v$. The velocity $v$ of the fractional element can be expressed as the sum of its translational part ($v^{(1)}$) and rotational motion ($\dot\alpha\hat{k} \times r_{1x}$), which is 
\begin{equation}
	\vec{v} = \vec{v}^{(1)} + \dot\alpha\hat{k} \times \vec{r}_{1x},
\end{equation}
    where $\vec{v}^{(1)}=v_{x}^{(1)}\hat{i}+v_{y}^{(1)}\hat{j}$, $\vec{r}_{1x}=r\cos\alpha \hat{i}+r\sin\alpha \hat{j}$, and
\begin{equation}\label{eqvelocity}\begin{split}
    	\vec{v} =& (v_{x}^{(1)}-\dot \alpha r\sin\alpha)\hat{i}+(v_{y}^{(1)}+\dot \alpha r\cos\alpha)\hat{j},\\
    	|\vec{v}| = & \sqrt{v_{x}^{(1)2}+v_{y}^{(1)2}+\dot\alpha^2r^2+2\dot\alpha r(v_{x}^{(1)}\sin\alpha-v_{y}^{(1)}\cos\alpha)},\\ 
    	\psi = & \tan^{-1}(\dfrac{v_{y}^{(1)}+\dot \alpha\cos\alpha}{v_{x}^{(1)}-\dot \alpha\sin\alpha}),\\
    	\vec{v}_n=&\vec{v}\sin(\psi-\alpha),\ \vec{v}_t=\vec{v}\cos(\psi-\alpha).    	
   \end{split} \end{equation}
Substituting the values of $v_n$ and $v_t$ from Eqn. \eqref{eqvelocity} into Eqn. \eqref{eqdragfractional}, the normal and tangential drag force acting on fractional element $X$ are obtained as
\begin{equation}\label{eqdrag} \begin{split}
	d\mathcal{D}_n = &\dfrac{1}{2}\rho C_{D,n} d [(v_{x}^{(1)}-\dot\alpha r\sin\alpha)\hat{i} +(v_{y}^{(1)}+\dot\alpha r\cos\alpha)\hat{j}][v_{x}^{(1)}\\ & \sin\alpha-v_{y}^{(1)}\cos \alpha-\dot\alpha r]^2(|\vec{v}(r)|)^{-1}dr,\\
	d\mathcal{D}_t =& \dfrac{1}{2}\rho C_{D,t}\pi d [(v_{x}^{(1)}-\dot\alpha r\sin\alpha)\hat{i} +(v_{y}^{(1)}+\dot\alpha r\cos\alpha)\hat{j}][v_{x}^{(1)}\\ &\cos\alpha-v_{y}^{(1)}\sin \alpha]^2(|\vec{v}(r)|)^{-1}dr.
\end{split}\end{equation}
The drag forces can be expressed in functional form as $(\mathcal{D}_n, \mathcal{D}_t)=f(\alpha,\dot\alpha,v_{x}^{(1)},v_{y}^{(1)},C_{D},d,l)$. These forces, acting over the entire rigid segment, are obtained by integrating Eqn. \eqref{eqdrag} along the segment length, \emph{i.e.}, from 0 to $l$.

 The moment generated due to normal drag force at node $i$ of Fig. \ref{figfreebody} is given in Eqn. \eqref{eqmomentdrag}. This expression is expanded below by substituting the drag force expression from Eqn. \eqref{eqdrag}, and is given as
\begin{equation*}\begin{split}
	\textstyle\sum M_{D,k}^{(i)} &=  \int_{o}^{l^{(i)}} \vec{r}\times d\vec{\mathcal{D}}_{n,k} dr,\\
	&=  \int_{o}^{l^{(i)}} (r\cos\alpha_k^{(i)}+r\sin\alpha_k^{(i)})\times (-\dfrac{1}{2}\rho C_{D,n} d^{(i)} \\ & [(v_{x,k}^{(i)} -\dot\alpha_k^{(i)} r  \sin\alpha_k^{(i)})\hat{i} +(v_{y,k}^{(i)}+\dot\alpha_k^{(i)} r\cos\alpha_k^{(i)})\hat{j}]\\ &[v_{x,k}^{(i)}\sin\alpha_k^{(i)}-v_{y,k}^{(i)} \cos \alpha_k^{(i)}-\dot\alpha_k^{(i)} r]^2(|\vec{v}_k^{(i)}(r)|)^{-1})dr, 
    \end{split}\end{equation*}
    or,
\begin{equation}\label{eqmomentdragintegral}\begin{split}
\textstyle\sum M_{D,k}^{(i)}	&=   -\frac{1}{2}\rho C_{D,n} d^{(i)} \int_{0}^{l^{(i)}} r (a_{y,k}^{(i)}\cos\alpha_k^{(i)}-a_{x,k}^{(i)}\sin\alpha_k^{(i)}\\ & -\dot{\alpha_k}^{(i)} r)^3  |\vec{v}_k(r)|^{-1} dr,
\end{split}\end{equation}
where \begin{equation*}\begin{split}|\vec{v}^{(i)}(r)|= &(v_{x}^{(i)2}+v_{y}^{(i)2}+\dot\alpha^{(i)2}r^2+2\dot\alpha^{(i)} r(v_{x}^{(i)}\sin\alpha^{(i)}-\\ & v_{y}^{(i)}\cos\alpha^{(i)}))^{1/2}.\end{split}\end{equation*}
Similarly, the moment generated at node $i$ in Fig. \ref{figfreebody} due to reaction force as given in Eqn. \eqref{eqmomentreaction} is expanded below by substituting the drag force expression from Eqn. \eqref{eqdrag}, and is given as
\begin{equation}\label{eqmomentreactionintegral}\begin{split}
	\textstyle\sum M_{rtn,k}^{(i)} = & -(m^{(i)} a_{x,CG,k}^{(i+1)} -\mathcal{D}_{n,x,k}^{(i+1)} - \mathcal{D}_{t,x,k}^{(i+1)}+ F_{x,k}^{(i+2)})\\ &(l^{(i)}\sin{\alpha}_{k}^{(i)}) +(m^{(i)} a_{y,CG,k}^{i+1} -\mathcal{D}_{n,y,k}^{(i+1)} - \mathcal{D}_{t,y,k}^{(i+1)} \\ & + F_{y,k}^{(i+2)})(l^{(i)}\cos{\alpha}_{k}^{(i)}).
\end{split}\end{equation}
The normal and tangential drag force components acting on $(i+1)^{th}$ segment can be expressed in functional form as   $(\mathcal{D}_{n,x,k}^{(i+1)},\mathcal{D}_{n,y,k}^{(i+1)})=f(k,\alpha^{(i)},\dot\alpha^{(i)},v_{x}^{(i)},v_{y}^{(i)},C_{D,n},d^{(i)},l^{(i)})$, and  $(\mathcal{D}_{t,x,k}^{(i+1)},\mathcal{D}_{t,y,k}^{(i+1)})=f(k,\alpha^{(i)},\dot\alpha^{(i)},v_{x}^{(i)},v_{y}^{(i)},C_{D,t},d^{(i)},l^{(i)})$ respectively. These drag force components are obtained by integrating Eqn. \eqref{eqdrag}, along the segment length, \emph{i.e.} 0 to $l^{(i+1)}$.

The moment balance equation at node $i$ of Fig. \ref{figfreebody} is obtained by substituting Eqns. \eqref{eqmomentdragintegral} and \eqref{eqmomentreactionintegral} into Eqn. \eqref{eqmomentexpanded}. It is expressed as
\begin{equation}\label{eqmomentfinal}\begin{split}
&-\frac{1}{2}\rho C_{D,n} d^{(i)} \int_{0}^{l^{(i)}} r (a_{y,k}^{(i)}\cos\alpha_k^{(i)}-a_{x,k}^{(i)}\sin\alpha_k^{(i)}-\dot{\alpha_k}^{(i)} r)^3  \\ &|\vec{v}_k(r)|^{-1} dr	+(m^{(i)} a_{x,CG,k}^{(i+1)} -\mathcal{D}_{n,x,k}^{(i+1)} - \mathcal{D}_{t,x,k}^{(i+1)} + 
F_{x,k}^{(i+2)})\\&(l^{(i)}\sin{\alpha}_{k}^{(i)}) +(m^{(i)}a_{y,CG,k}^{(i+1)} 
 -\mathcal{D}_{n,y,k}^{(i+1)} - \mathcal{D}_{t,y,k}^{(i+1)} + 
F_{y,k}^{(i+2)}) \\&(l^{(i)}\cos{\alpha}_{k}^{(i)})=
I^{(i)}\ddot\alpha_k^{(i)} + \frac{m^{(i)} l^{(i)}}{2}(a_{y,k}^{(i)} \cos\alpha_k^{(i)} - \\&a_{x,k}^{(i)} \sin\alpha_k^{(i)} )+\mathcal{C}_{damp}\dot\alpha_k^{(i)},
\end{split}\end{equation}
It can be inferred from Eqn. \eqref{eqmomentfinal} that the moment balance equation at $i^{th}$ node ($MBE^{(i)}$) is a function of azimuth, position, velocity, acceleration, and physical parameters of $i^{th}$ rigid segment. This can be expressed as $MBE_i=f(k,\alpha^{(i)},\dot\alpha^{(i)},\ddot\alpha^{(i)},x^{(i)},y^{(i)},v_{x}^{(i)},v_{y}^{(i)},a_{x}^{(i)},a_{y}^{(i)},l^{(i)},m^{(i)},d^{(i)}).$
The azimuth $\alpha^{(i)}$ for $i^{th}$ segment is obtained by solving the $2^{nd}$ order ordinary differential Eqn. \eqref{eqmomentfinal} defined by $MBE^{(i)}$. The computed value of $\alpha^{(i)}$ is then used in kinematic equations provided below to determine the position, velocity and acceleration at $(i+1)^{th}$ node of towed segment, such that 
\begin{equation} \label{eqkinematics} \begin{split}
		& x^{(i+1)}_k = x^{(i)}_k + l^{(i)} \cos\alpha_k^{(i)}, \ y^{(i+1)}_k = y^{(i)}_k + l^{(i)} \sin\alpha_k^{(i)},\\
		& v_{x,k}^{(i+1)} = v_{x,k}^{(i)}-l^{(i)}\dot{\alpha_k}^{(i)}\sin\alpha_k^{(i)}, \\ & v_{y,k}^{(i+1)} =v_{y,k}^{(i)}+ l^{(i)}\dot{\alpha_k}^{(i)}\cos\alpha_k^{(i)}, \\
		& a_{x,k}^{(i+1)}= a_{x,k}^{(i)}-l^{(i)}\ddot{\alpha_k}^{(i)}\sin\alpha_k^{(i)}-l^{(i)}\dot{\alpha_k}^{(i)2}\cos\alpha_k^{(i)}, \\ & 
		a_{y,k}^{(i+1)}=  a_{y,k}^{(i)}+l^{(i)}\ddot{\alpha_k}^{(i)}\cos\alpha_k^{(i)}-l^{(i)}\dot{\alpha_k}^{(i)2}\sin\alpha_k^{(i)}.
\end{split}\end{equation}
The position ($x^{(1)},y^{(1)}$), velocity ($v_{x}^{(1)},v_{y}^{(1)}$) and acceleration ($a_{x}^{(1)},a_{y}^{(1)}$) at the first node of the lumped mass model for TCSAS, as shown in Fig. \ref{figfreebody}, are identical to the corresponding values at the stern of the tow-ship.
Therefore, these values are assumed to be known at all time instants from the onboard inertial navigation system of the tow-ship \cite{SINGH2024104774}.
The moment balance equation at the first node is, $MBE_k^{(1)}=f(k,\alpha^{(1)},\dot\alpha^{(1)},\ddot\alpha^{(1)},x^{(1)},y^{(1)},v_{x}^{(1)},v_{y}^{(1)},a_{x}^{(1)},a_{y}^{(1)},l^{(1)},m^{(1)}, d^{(1)})$, these function parameters are utilized in Eqn. \eqref{eqmomentfinal} to obtain a second order ODE of azimuth $\alpha^{(1)}$.
This ODE is solved to get the value of $\alpha^{(1)}$, which is further passed through kinematic Eqn. \eqref{eqkinematics} to get the position ($x^{(2)},y^{(2)}$), velocity ($v_{x}^{(2)},v_{y}^{(2)}$) and acceleration ($a_{x}^{(2)},a_{y}^{(2)}$) at second node of lumped mass model for TCSAS. 
In a similar manner, the moment balance equation at node 2 is 
\begin{equation*}
\begin{split}
MBE^{(2)}=&f(k,\alpha^{(2)},\dot\alpha^{(2)},\ddot\alpha^{(2)},x^{(2)},y^{(2)},v_{x}^{(2)},v_{y}^{(2)},a_{x}^{(2)},\\
&a_{y}^{(2)},l^{(2)},m^{(2)},d^{(2)}),     
\end{split}
 \end{equation*} which is a second order ODE of $\alpha^{(2)}$. It is solved to get the azimuth $\alpha^{(2)}$, which is further passed through kinematic Eqn. \eqref{eqkinematics} to get the position, velocity and acceleration at node 3.
This sequential process is repeated starting from the top node $(n=1)$ to $(n-1)^{th}$ node of lumped mass model to determine the azimuth, position, velocity, and acceleration at each node of TCSAS.

The moment balance equation for $n^{th}$ node at $k^{th}$ time instant for TCSAS, as shown in Fig. \ref{figfreebody}, doesn't include the moment due to reaction force, since the other end is free. It is expressed as
\begin{equation}\label{eqmomentsensorarray}\begin{split}
		& -\frac{1}{2}\rho C_{D,n} d^{(n)} \int_{0}^{l^{(n)}} r (a_{y,k}^{(n)}\cos\alpha_k^{(n)}-a_{x,k}^{(n)}\sin\alpha_k^{(n)}-\dot{\alpha_k}^{(n)} r)^3 \\& |\vec{v}_k(r)|^{-1} dr 	=
		I^{(n)}\ddot\alpha_k^{(n)} + \frac{m^{(n)} l^{(n)}}{2}(a_{y,k}^{(n)} \cos\alpha_k^{(n)} - \\ &a_{x,k}^{(n)} \sin\alpha_k^{(n)} )+\mathcal{C}_{damp}\dot\alpha_k^{(n)}.
\end{split}\end{equation}
The solution for $2^{nd}$ order ODE \eqref{eqmomentsensorarray} gives the azimuth orientation $\alpha^{(n)}$ of $n^{th}$ node. This azimuth is then used in kinematic Eqn. \eqref{eqkinematics} to compute the position of CG of towed sensor-array in $xy$ plane.

The above mentioned procedure is systematically followed at each time instant to get the azimuth and location of each lumped mass nodes of TCSAS in $xy$ plane. 
\subsection{Quasi-static equilibrium condition} \label{subsectionquasi}
\begin{figure*}[ht]
	\centering
    \includegraphics[width=\textwidth]{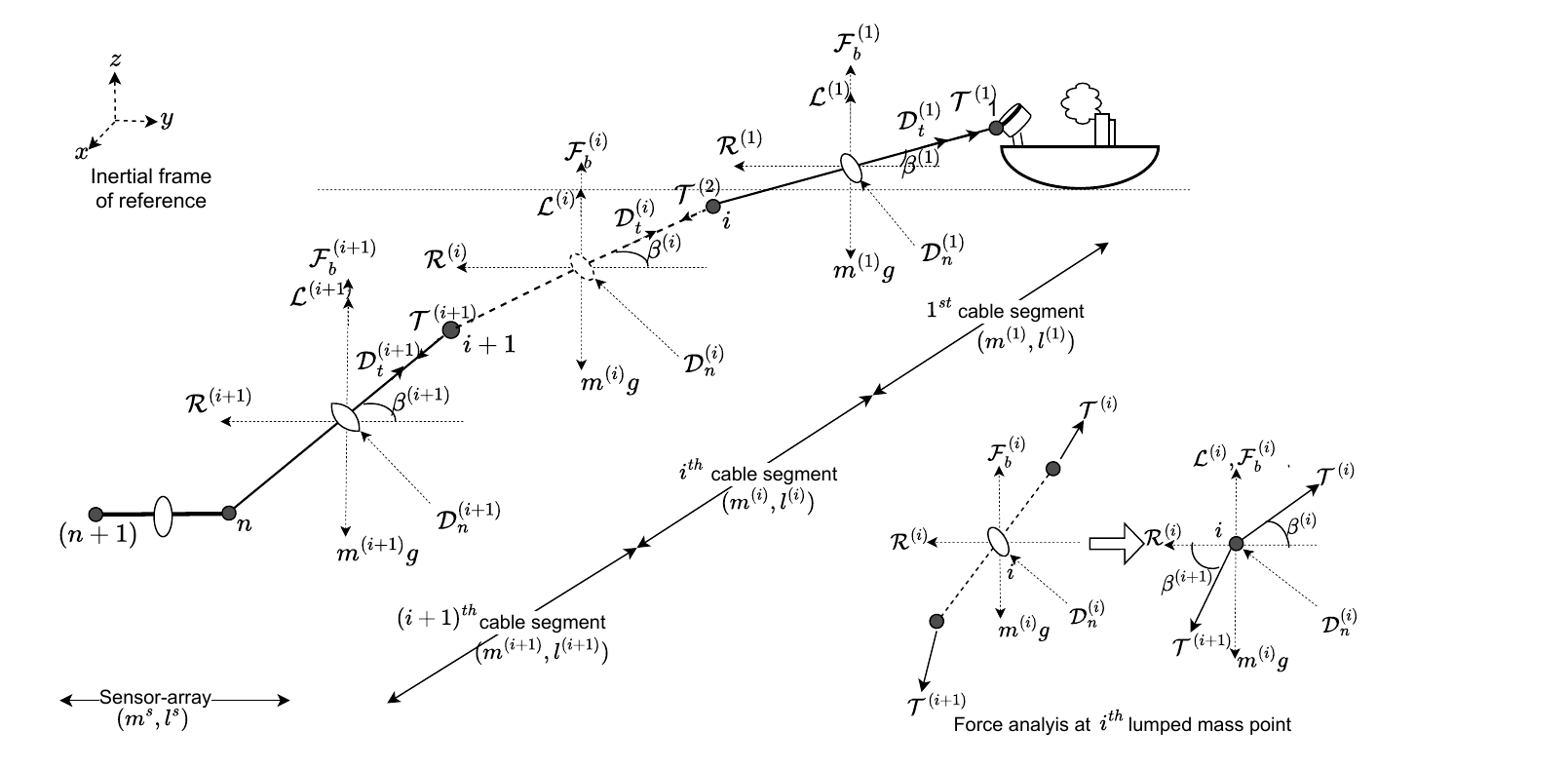} 
	\caption{Lumped mass model of TCSAS for quasi-static equilibrium analysis in z-plane.}
	\label{figquasi}
\end{figure*}

The towed cable is modelled flexible using $(n-1)$ discretized rigid segment interconnected at ($n-1$) nodes. The sensor-array is connected to $n^{th}$ node of cable using a revolute joint.
The diagrammatic representation of TCSAS in $z$-plane is shown in Fig. \ref{figquasi} \cite{liu2023study}. The configuration of TCSAS in $z$-plane is determined by the elevation angle $\beta$ of each rigid segment and the depths of the rigid segments \emph{w.r.t.} sea surface.
The elevation angle ($\beta$) is the angle between the rigid links and the horizontal plane, or the angle between tension acting on the rigid links and the horizontal plane.
The depth and elevation angle of rigid segments is obtained by performing the force balance under the quasi-static equilibrium condition at the nodes \cite{liu2023study,sanders1982three}.
The forces due to tension, gravity, buoyancy, hydrodynamics (lift and resistance) are considered to be acting on the lumped mass point. The lumped mass point is considered to be at the centre of rigid segment as shown in Fig. \ref{figquasi}.
The configuration for each rigid segment in $z$-plane is determined by using force balance condition at the lumped mass point. The resulting equations are solved to get the tension acting on each segment, elevation angle \emph{w.r.t.} horizontal plane and depth of the segment.
The buoyancy and force due to gravity are calculated based on physical properties of rigid segments, and the hydrodynamic forces are calculated using the Morison equation \cite[ch.9]{nakayama2018introduction}. 
 
The equations derived from force balance condition at $i^{th}$ lumped mass point at $k^{th}$ time instant from Fig. \ref{figquasi} is given as
\begin{equation}\label{eqforcebalance}\begin{split}
	& \mathcal{T}_k^{(i)} \sin\beta_k^{(i)}+\mathcal{F}_{b,k}^{(i)}+\mathcal{L}_k^{(i)}=m^{(i)} g+\mathcal{T}_k^{(i+1)} \sin\beta_k^{(i+1)},\\
	& \mathcal{T}_k^{(i)}\cos\beta_k^{(i)}= \mathcal{T}_k^{(i+1)}\cos\beta_k^{(i+1)}+\mathcal{R}_k^{(i)}.
\end{split}\end{equation} 
  $\beta^{(i)}$ is elevation angle of $i^{th}$ segment \emph{w.r.t.} horizontal $xy$ plane, and $\mathcal{F}_{b}^{(i)}$ is the buoyant force on $i^{th}$ segment. $\mathcal{T}^{(i)}$ and $\mathcal{T}^{(i+1)}$ is the tension acting on $i^{th}$ and $(i+1)^{th}$ segment, respectively. $\mathcal{L}^{(i)}$ and $\mathcal{R}^{(i)}$ are lift and resistance force due to hydrodynamics on $i^{th}$ segment, respectively, which are expressed as  
\begin{equation}\label{eqliftresistance}\begin{split}
	&\mathcal{L}_k^{(i)} = \mathcal{D}_{n,k}^{(i)}\cos\beta_k^{(i)}+\mathcal{D}_{t,k}^{(i)}\sin\beta_k^{(i)},\\
	&\mathcal{R}_k^{(i)}=  \mathcal{D}_{n,k}^{(i)}\sin\beta_k^{(i)}-\mathcal{D}_{t,k}^{(i)}\cos\beta_k^{(i)}.
\end{split}\end{equation}
 $\mathcal{D}_{n}^{(i)}$ and $\mathcal{D}_{t}^{(i)}$ are normal and tangential drag force acting on $i^{th}$ segment, respectively. They are expressed using the Morison formula as \cite[ch.9]{nakayama2018introduction}
\begin{equation}\label{eqdragquasi}\begin{split}
\mathcal{D}_{n}^{(i)} = & -\dfrac{1}{2} \rho C_{D,n} A^{(i)} \vec{v}_{n}^{(i)} |\vec{v}_{n}^{(i)}|,\\
\mathcal{D}_{t}^{(i)} = & -\dfrac{1}{2}\rho C_{D,t} \pi A^{(i)} \vec{v}_{t}^{(i)} |\vec{v}_{t}^{(i)}|,
\end{split}\end{equation}
where $A^{(i)}$ is the surface area of the $i^{th}$ segment. $v_{n}^{(i)}$ and $v_{t}^{(i)}$ are the normal and tangential velocity component of $i^{th}$ segment, which are related to velocity $v^{(i)}$ in terms of expression
\begin{equation}\label{eqvelocityquasi}
	|v_{n}^{(i)}| = |v^{(i)}|\sin\beta^{(i)},\ |v_{t}^{(i)}| = |v^{(i)}|\cos\beta^{(i)}.
\end{equation}
 $v^{(i)}$ is deduced from moment balance condition at node $i$ using the kinematic Eqn. \eqref{eqkinematics}.

The one end of sensor-array is connected to $n^{th}$ node of cable via a revolute joint , such that its elevation angle is fixed at $\beta^{(n)}=0^o$. The other end of the sensor array is free, resulting in zero tension at that point, \emph{i.e.} $\mathcal{T}^{(n+1)}=0$.
The force balance equations for the sensor array at the $n^{th}$ lumped mass point are derived using Eqns. \eqref{eqforcebalance}-\eqref{eqdragquasi}, and are expressed as
\begin{equation}\begin{split}
	\mathcal{F}_{b,k}^{(n)}= m_sg, \  \mathcal{T}_k^{(n)}=-\mathcal{D}_{t,k}^{(n)}.
\end{split}\end{equation}
The tension force  $\mathcal{T}^{(n)}$ and elevation angle $\beta^n$ of sensor-array segment are subsequently utilised in force balance Eqns. \eqref{eqforcebalance}-\eqref{eqdragquasi} at the $(n-1)^{th}$ lumped mass point to compute its elevation angle and tension acting on $(n-1)^{th}$ segment.
In this way, the force balance equations are solved starting from bottom-most segment \emph{i.e.} sensor-array and proceeding sequentially upward toward the top segment, using Eqns. \eqref{eqforcebalance}-\eqref{eqvelocityquasi}, to compute the tension force and elevation angle of each segment.

The depth ($h$) attained by lumped mass modelled rigid segments is given as
\begin{equation}\label{eqdepth}
	h_k^{(i)}= \begin{cases}
		\dfrac{1}{2}l^{(i)} \sin\beta_k^{(i)},\ \text{for}\ i=1,\\
		 \textstyle \sum_{m=1}^{i-1}l^{(m)}\sin\beta_k^{(m)}+\dfrac{1}{2}l^{(i)}\sin\beta_k^{(i)}, \ i=2\ldots n-1,\\
		 \textstyle\sum_{m=1}^{n-1}l^{(m)}\sin\beta_k^{(m)}+\dfrac{1}{2}l^{(i)} \sin\beta_k^{(i)}, \ \text{for} \ i=n.
	\end{cases}
\end{equation}
The azimuth, position, velocity and acceleration of each interconnected segment in $xy$ plane are obtained using the moment balance condition, as described in Subsection \ref{subsectionmoment}. The elevation, depth, and tension acting for each rigid segment are computed using the quasi-static equilibrium condition developed in Subsection \ref{subsectionquasi}. The equations derived from both conditions are solved simultaneously to localize each interconnected segment in $3D$ space. The resulting sensor-array position is then fed to a state estimation algorithm running on the onboard computer to estimate the target state. 

\section{Maximum correntropy Filtering}
Conventional nonlinear Kalman filters use the MMSE criterion, which works well when the measurement noise is Gaussian because it captures up to second-order error statistics \cite{liu2007correntropy}. However, when the noise is impulsive or non-Gaussian, MMSE becomes less effective. In such cases, an information-theoretic learning approach called correntropy can be used \cite{liu2007correntropy}. Correntropy captures higher-order even moments of error, making it more suitable for handling non-Gaussian noise \cite{chen2017maximum}. A filtering algorithm based on the maximum correntropy criterion (MCC) can therefore improve robustness against non-Gaussian measurement noise \cite{hou2017maximum}.

\subsection{Correntropy}
Correntropy is a measure that quantifies the similarity between two random variables \cite{chen2017maximum}. For two random variables $x$ and $y$, having the joint pdf $f(x,y)$, correntropy is defined as
\begin{equation}
	V(x,y) = \mathbb{E}[K(x,y)] = \int_{x} \int_{y} K(x,y)f(x,y)dxdy,
\end{equation} 
where $K(\cdot)$ is the kernel function \cite{principe2010information}, which evaluates the similarity between sample pairs. We consider the kernel function to be Gaussian, which is given as
\begin{equation}
	K(x,y)=G_{\sigma}(e)=exp\dfrac{-(x-y)^2}{2\sigma^2},
\end{equation}
where $e=x-y$ is the error and $\sigma$ is kernel bandwidth. 

In practical applications, the joint probability density function is often unavailable. Instead, correntropy is computed using discrete sample data $\{ (x_i,y_i) \}_{i=1}^m$, resulting in the sample mean estimate. The correntropy is expressed as
\begin{equation} \label{eqnkernel}
	\hat{V}(x,y) = \dfrac{1}{n}\sum_{i=1}^{m} G_{\sigma}(e_i), 
\end{equation} 
where $n$ is the number of sample points. The Taylor series expansion of the kernel function in Eqn. \eqref{eqnkernel} yields 
\begin{equation}
	\hat{V}(x,y) = \dfrac{1}{\sqrt{2\pi}\sigma}\sum_{m=0}^{\infty} \dfrac{-1^m}{2^mm!}\mathbb{E}\dfrac{(x-y)^m}{\sigma^{2m}}.
\end{equation}
The above expression shows that correntropy accounts for a weighted combination of even-order moments of the error ($e=x-y$), with the kernel bandwidth $\sigma$ acting as a scaling parameter.
Unlike MMSE, which captures only second-order moments, correntropy incorporates higher-order moments, making it inherently robust against impulsive and non-Gaussian noise \cite{liu2007correntropy}. As $\sigma$ increases significantly, higher-order contributions diminish, and correntropy approaches the MMSE criterion. 
\subsection{General framework for maximum correntropy Kalman filter}
The BOT is a nonlinear filtering problem with the equation describing the process and measurement model is recalled here from Eqn. \eqref{eqnprocessmodel} and \eqref{eqnmeasurementmodel} as \cite{radhakrishnan2018new}
\begin{equation}\begin{split}\label{eqmodel}
		X_k = & FX_{k-1}+\mu_{k-1},\\
		Y_k = & \Phi(X_k)+r_k,
\end{split}\end{equation}
where $X_k\in  \mathbb{R}^n$ and $Y_k\in  \mathbb{R}^m$ are the state and measurement vector at time $k$. 
$\Phi(\cdot): \mathbb{R}^n \rightarrow \mathbb{R}^m$ is the measurement model. Process noise is Gaussian distributed represented as $\mu_{k-1}\sim\mathbb{N}(0,Q_{k-1})$.
The bearing data is collected from a towed sensor array and is corrupted with non-Gaussian noise $r_k$, which is modelled as a weighted sum of Gaussian with equivalent covariance $R_k$ \cite{saha2023robust, chen2017maximum, liu2007correntropy}. 
The system states are estimated in two steps, namely (i) time update, and (ii) measurement update.
\subsubsection{Time update} 
The time update stage uses the state transition matrix to predict the state estimate. The predicted pdf or the prior pdf is obtained using the Chapman-Kolmogorov equation \cite{kumar2023new}, expressed as 
\begin{equation}
	p(X_k|Y_{1:k-1}) = \int p(X_k|X_{k-1}) p(X_{k-1}|Y_{1:k-1}) dX_{k-1},
\end{equation}
where $p(X_k|X_{k-1})$ represents the state transition pdf, and $p(X_{k-1}|Y_{1:k-1})$ is the posterior density from the previous time step.
The prior state estimate $(\hat{X}_{k|k-1})$ is expected value of prior pdf, given as \cite{SINGH2024104774} 
\begin{equation}\label{eqxpredict}\begin{split}
	\hat{X}_{k|k-1} = & \mathbb{E}[p(X_k|Y_{1:k-1})],\\
    = & \int FX_{k-1} p(X_{k-1}|Y_{1:k-1}) dX_{k-1},\\
    = & \int FX_{k-1} \mathcal{N}(X_{k-1}; \hat{X}_{k-1|k-1}, P_{k-1|k-1}) dX_{k-1},\\
     = & F\hat{X}_{k-1|k-1}.
\end{split}\end{equation}
The prior error covariance is calculated as \cite{kumar2023new, SINGH2024104774}
\begin{equation}\label{eqPpredict}
	\begin{split}
		P_{k|k-1} = & \mathbb{E}[(X_k - \hat{X}_{k|k-1})(X_k - \hat{X}_{k|k-1})^\top | Y_{1:k-1}] \\
        = & FP_{k-1|k-1}F^\top +Q_k,
	\end{split}
\end{equation}
where $\top$ denotes transpose.
It is important to note that the time update step, which relies solely on the process model, is not directly influenced by the non-Gaussian nature of the measurement noise \cite{chen2017maximum}. 
\subsubsection{Measurement update}
A cost function is formulated using MCC to handle non-Gaussian measurement noise, and measurement update equations are derived by maximizing this cost function.
Firstly, the process and measurement model from Eqn. \eqref{eqmodel} are augmented as shown below 
\begin{equation}\label{eqaugment}
		\begin{bmatrix} 
		\hat{X}_{k|k-1} \\ Y_k	\end{bmatrix} = \begin{bmatrix} X_k \\ \Phi(X_k)
	\end{bmatrix} + \nu_k,
\end{equation}
where $\nu_k = \begin{bmatrix}  -(X_k -\hat{X}_{k|k-1}) \\ r_k \end{bmatrix}$, with covariance
\begin{equation} \label{eq22n} \begin{split} 
		\mathbb{E}(\nu_k\nu_{k,\top}) & = \begin{bmatrix} P_{k|k-1} & 0 \\ 0 & R_k \end{bmatrix} = \begin{bmatrix}  S_{k|k-1}S_{k|k-1,\top} & 0 \\ 0 & 	S_{R,k}S_{R,k}^\top \end{bmatrix}\\ & =  B_kB_{k}^\top,
	\end{split} 
\end{equation} 
\begin{equation}\label{eqchol}
	S_{k|k-1}=chol(P_{k|k-1}),\ S_{R,k}=chol(R_k),
\end{equation}
where $B_k=diag(S_{k|k-1},S_{R,k})$. Left multiplying both side of Eqn. \eqref{eqaugment} by $B_k^{-1}$ transforms the recursive model in Eqn. \eqref{eqaugment} to a non-linear regression model, given as
\begin{equation} 
	B_k^{-1} 	\begin{bmatrix} \hat{X}_{k|k-1} \\ Y_k	\end{bmatrix} =  B_k^{-1} \begin{bmatrix} X_k \\ \Phi({X}_{k})	\end{bmatrix} +  \mathcal{E}_k, 
\end{equation}
where the weighted error matrix ($\mathcal{E}_k$) given by
\begin{equation} \label{eq23}
	\mathcal{E}_k =  B_k^{-1} \nu_k = \begin{bmatrix}
		-S_{k|k-1}^{-1} (X_k-\hat{X}_{k|k-1}) \\ S_{R,k}^{-1} (Y_k-\Phi({X}_{k}))  \end{bmatrix}.
\end{equation}
Since, $\mathbb{E}[\mathcal{E}_k\mathcal{E}_{k}^\top]=I$, the weighted error matrix $\mathcal{E}_k$ is white.

Now, the cost function is defined as an exponential sum of squared error vectors weighted by a kernel bandwidth, given as
\begin{equation} \label{eq24}
	\mathclap{J}_{MCC}(X_k)  = \frac{1}{n+m} \sum_{l=1}^{n+m} exp ( \frac{-e_{l,k}^{2}}{2\sigma^2}),
\end{equation}
where ${e_{l,k}}$ is the $l^{th}$ element of weighted error matrix $\mathcal{E}_{k}$ at $k^{th}$ instant. From (\ref{eq23}), $e_{l,k}$ represents the weighted state prediction error for $l=1:n$ and weighted measurement error from $l=n+1:n+m$, such that
\begin{equation}\label{eqekj}\begin{split}
		e_{l,k} = & -S_{k|k-1}^{-1} (X_k-\hat{X}_{k|k-1}), \ l= 1,2 \cdots n,\\
		e_{l,k} = & S_{R,k}^{-1} (Y_k-\Phi({X}_{k})), \ l=n+1 \cdots n+m.
	\end{split}     
\end{equation}
The objective is to find the posterior state estimate ($\hat{X}_{k|k}$) using the measurement data $Y_{1:k}$, which can be obtained by maximizing the cost function \eqref{eq24} given as
\begin{equation} \label{eq25}
	\hat{X}_{k|k} = \arg \; \max_{X_k} \ J_{MCC}(X_k).
\end{equation}
Now, a weighting parameter called correntropy matrix $\mathit{\Pi}_k$, is defined as \cite{liu2018maximum,chen2017maximum,liu2016extended,liu2016maximum,liu2019linear,liu2007correntropy,singh2010closed,zhao2022robust}
\begin{equation}
	\mathit{\Pi}_k = \begin{bmatrix}
		\Pi_{P,k} & 0 \\ 0 & \Pi_{R,k}
	\end{bmatrix}, \end{equation}
where $\mathit{\Pi}_{j,k}=\exp(-\frac{e_{l,k}^2}{2\sigma^2})$ and 
\begin{equation} \label{eq29} \begin{split}
		\Pi_{P,k} & = diag (\begin{bmatrix} \mathit{\Pi}_{1,k} & \mathit{\Pi}_{2,k} & \cdots & \mathit{\Pi}_{n,k}\end{bmatrix}_{1\times n}),\\
		\Pi_{R,k} & = diag (\begin{bmatrix} \mathit{\Pi}_{n+1,k} & \mathit{\Pi}_{n+2,k} & \cdots & \mathit{\Pi}_{n+m,k} \end{bmatrix}_{1\times m}).
	\end{split} 
\end{equation}
So, the modified prior error covariance and modified measurement noise covariance are \cite{liu2018maximum,chen2017maximum}
\begin{equation} \label{eqp} 
	\bar{P}_{k|k-1} = S_{k|k-1}\Pi_{P,k}^{-1}S_{k|k-1}^\top,
\end{equation}
and 
\begin{equation} \label{eqr} 
	\bar{R}_{k} = S_{R,k}\Pi_{R,k}^{-1}S_{R,k}^\top,
\end{equation}
respectively.

To perform the measurement update for $k^{th}$ step, the modified prior error covariance $\bar{P}_{k|k-1}$, and modified measurement noise covariance $\bar{R}_{k}$ are to be calculated, which are the functions of error vector $e_{l,k}$ as mentioned in Eqn. \eqref{eqekj}.
The calculation of this error vector demands the true value of state $X_k$, which is unknown.
To overcome this problem, a fixed point iteration (FPI) method is employed, which iteratively calculates posterior state estimate $\hat{X}_{k|k}^i$ for a time $k$. Here, the subscript $i$ denotes a fixed point iteration step at time instant $k$.
This FPI method is initiated by assigning the prior state estimate as the truth value, \emph{i.e.} $\hat{X}_{0|0,k}=\hat{X}_{k|k-1}$ \cite{chen2017maximum,chen2019minimum,liu2016extended} and then calculate $e_{l,k}^{i}$, $\mathit{\Pi}_{k}^{i}$, $\bar{P}_{k|k-1}^{i}$, and $\bar{R}_{k}^{i}$ using \eqref{eqekj}, \eqref{eq29}, \eqref{eqp}, and \eqref{eqr}, respectively. 
 The Posterior state estimate $\hat{X}_{k|k}^{i+1}$ is updated iteratively through the measurement update process, which is described below.

The posterior state pdf is obtained using the Bayes' rule \cite{kumar2023new}, given as
\begin{equation}
	p(X_k|Y_k)  \propto p(Y_k|X_k)p(X_k|Y_{1:k-1}),
\end{equation}
where $p(Y_k|X_k)\sim\mathcal{N}(Y_k;\hat{Y}_{k|k-1},P_{YY,k|k-1})$ is the likelihood approximated as Gaussian. The predicted measurement mean is 
\begin{equation} \label{eqy}
	\hat{Y}_{k|k-1} = \int \Phi(X_k)\mathcal{N}(X_k;\hat{X}_{k|k-1},\bar{P}_{k|k-1})dX_k,
\end{equation}
and the $i^{th}$ iterated innovation error covariance matrix is
\begin{equation}\label{eqpy} \begin{split}
		P_{YY,k|k-1}^i = & \int \Phi(X_k)\Phi(X_k)^\top \mathcal{N}(X_k;\hat{X}_{k|k-1},\bar{P}_{k|k-1})dX_k -\\ & \hat{Y}_{k|k-1}\hat{Y}_{k|k-1}^\top+\bar{R}_k^i.
\end{split}\end{equation}
The cross-covariance matrix is independent of modified error covariance $(\bar{P}_{k|k-1})$ and modified measurement noise covariance ($\bar{R}_k$) term, such that it is calculated once before the execution of FPI. It is given as
\begin{equation}\label{eqpxy} \begin{split}
		P_{XY,k|k-1} = & \int X_k \Phi(X_k)^\top \mathcal{N}(X_k;\hat{X}_{k|k-1},\bar{P}_{k|k-1}) dX_k-  \\ & \hat{X}_{k|k-1} \hat{Y}_{k|k-1}^\top.
\end{split}\end{equation}
The posterior state estimate at (${i+1})^{th}$ iteration is
\begin{equation}\label{eqxkk}
	\hat{X}_{k|k}^{i+1} =  \hat{X}_{k|k-1}^i+K_k^i(Y_k-\hat{Y}_{k|k-1}),
\end{equation} 
where $K_k^i =  P_{XY,k|k-1}(P_{YY,k|k-1}^i)^{-1}$ is the iterated Kalman gain.

The posterior state estimate $(\hat{X}_{k|k}^{i+1})$ is updated iteratively using the measurement update process.  The iteration process is continued till the relative difference between consecutive iterations for the posterior estimated states is less than a pre-defined threshold value $\eta$, such that $\frac{||\hat{X}_{k|k}^{i+1} - \hat{X}_{k|k}^{i}||}{||\hat{X}_{k|k}^{i}||} \leq \eta$.

After convergence of the posterior state estimate, the posterior error covariance is computed using the last step iterated value of Kalman gain and innovation error covariance, given as
\begin{equation}\label{eqpkk}
	P_{k|k} =  \bar{P}_{k|k-1}-K_k P_{YY,k|k-1} K_k^\top.
\end{equation} 
The fixed point iterative approach ensures accurate computation of the posterior state and its covariance, despite the unavailability of the true state. The FPI algorithm is provided in Algorithm \ref{fpialgorithm}.
\begin{algorithm}
	\caption{Fixed point iteration (FPI)}
	\label{fpialgorithm}
	$[\hat{X}_{k|k}]: = FPI [\hat{X}_{k|k-1}, Y_k]$
	\begin{itemize}
		\item Calculate $\hat{X}_{k+1|k}$ using \eqref{eqxpredict} and initialize $\hat{X}_{k|k}^i= \hat{X}_{k+1|k}$.
		\item \textbf{for} $i=1:i_{max}$
		\begin{itemize}
			\item Calculate $e_{l,k}^{i}$, $\Pi_{p,k}^{i}$ and $\Pi_{R,k}^{i}$  using \eqref{eqekj}, \eqref{eq29}.
			\item Evaluate $\bar{P}_{k|k-1}^i$ and $\bar{R}_k^i$ using \eqref{eqp} and \eqref{eqr}.
			\item Calculate $P_{YY,k|k-1}^i$ using \eqref{eqpy}.
			\item Compute $\hat{X}_{k|k}^{i+1}$ using \eqref{eqxkk}. 
			\item \textbf{if} $\frac{||\hat{X}_{k|k}^{i+1} - \hat{X}_{k|k}^{i}||}{||\hat{X}_{k|k}^{i}||} \leq \eta$
			\begin{itemize}
				\item[-] $\hat{X}_{k|k} = \hat{X}_{k|k}^{i+1}$.
				\item [-] break
			\end{itemize}
			\textbf{else} $i = i + 1$.
		\end{itemize}
		\item \textbf{end for}       
	\end{itemize}
\end{algorithm}

The maximum correntropy filtering in the Bayesian framework presented above requires the evaluation of the integral of the form $\int (function \times Gaussian \ pdf)$ as can be seen in Eqns. \eqref{eqy}, \eqref{eqpy}, \eqref{eqpxy}. The integrals of such kind are computationally intractable for nonlinear functions, and their closed form solution does not exist. 
One of the first filtering approaches to address non-Gaussian measurement noise for nonlinear systems is introduced in MCEKF \cite{liu2016extended}. Here, the nonlinear function is linearised around the previous estimated states using a first-order Taylor series expansion.
Later, various deterministic sampling point filtering approach based on maximum correntropy criterion are proposed, such as MCUKF \cite{hou2018maximum}, MCCKF \cite{liu2018maximum}, and MCGHF \cite{qin2017maximum}. These filters tackled non-Gaussian sensor noise using MCC based cost function, and they approximated the intractable integrals by using predefined sample points and their associated weights.
The MCUKF utilizes the unscented transform to generate deterministic points \cite{hou2018maximum}, MCCKF applies a third-degree spherical-radial cubature rule to compute cubature points \cite{liu2018maximum}, and MCGHF employs Gauss-Hermite quadrature rule for integration \cite{qin2017maximum}. MC based nonlinear filters based on polynomial approximation for nonlinear function is proposed as MCPCKF \cite{SINGH2024104774}. It uses Hermite polynomial chaos expansion to model the stochastic process using Gaussian random variables \cite{SINGH2024104774, kumar2023polynomial}. 


For MCC-based filters, the kernel bandwidth is a crucial parameter, and making its appropriate selection is essential for achieving accurate state estimation \cite{cinar2012hidden, hou2017maximum, fakoorian2019maximum, saha2023robust}. In the next subsection, a kernel bandwidth selection technique is proposed.

\subsection{Selection of Kernel Bandwidth}
The state estimation accuracy of MC-based estimators critically depends on the choice of kernel bandwidth. Using a fixed kernel size throughout all time steps is inadequate because if the kernel is too large, the estimator behaves like a simple linear regressor and fails to account for measurement outliers. Conversely, if it is too small, every data point is treated as unique, impairing the estimator’s ability to generalize to new observations. Tuning the kernel bandwidth at each time step is therefore essential for obtaining accurate and reliable estimates \cite{cinar2012hidden,hou2017maximum}. 
Although several adaptive strategies have been proposed in the literature. Such as \cite{fakoorian2019maximum} calculates kernel size as the sum of the weighted innovation term and the weighted covariance of the estimation error at each step time, given as 
\begin{equation*}
\sigma_k=(||Y_k-\hat{Y}_{k|k-1}||_{R_{k(-1)}}+\Phi(\hat{X}_{k|k-1})P_{k|k-1}\Phi^\top(\hat{X}_{k|k-1}))^{-1}.
\end{equation*}
In \cite{cinar2012hidden}, the kernel size is heuristically chosen at each time instant to be the $2^{nd}$ norm of the innovation, given as $\sigma_k=||Y_k-\Phi(\hat{X}_{k|k-1})||$.
\cite{hou2017maximum} selects the kernel size as the Mahalanobis distance of the innovation given as 
\begin{equation*}
\sigma_k=||(Y_k-\Phi(\hat{X}_{k|k-1}))^\top R_{k,-1}(Y_k-\Phi(\hat{X}_{k|k-1}))||.
\end{equation*}
 The existing literature is typically problem-specific and does not guarantee an optimal solution across different scenarios.

To overcome this, we propose a cost function that selects the kernel bandwidth by minimizing the trace of the posterior error covariance matrix $P_{k|k}$.
The idea is to find the kernel bandwidth that leads to the most confident estimate, quantified by the smallest total variance of the state estimate.
The cost function is defined as
\begin{equation}
    J_{kernel}(\sigma_k) = Tr(P_{k|k}),
\end{equation}
where $Tr(\cdot)$ is the trace of a matrix.
The optimal kernel bandwidth is obtained by
\begin{equation} \label{eqnoptimalkernel}
    \sigma_{k}^* = \arg \; \min_{\sigma_k} \ J_{kernel}(\sigma_k).
\end{equation}
By evaluating this cost over a predefined range of $\sigma_k$, the kernel bandwidth that yields the minimum posterior uncertainty is selected, and the corresponding state estimate is taken as the final output of the estimator.

The 3D position of the sensor array’s center of gravity (CG) is determined using the dynamic modeling of the TCSAS, as detailed in Section \ref{sectiondynamicmodel}. Thus, the target state estimation during BOT, in the presence of non-Gaussian sensor noise and bearing data collected from the dynamically modelled towed sensor-array, is performed using MCC-based Kalman filter with towed array dynamics (MCC-TAD). The implementation steps for MCC-TAD are outlined in Algorithm \ref{algoBOT}.
\begin{algorithm} 
	\caption{ MCC-TAD}
	\label{algoBOT}
    $[\hat{X}_{k|k},P_{K|k}]:=\text{MCC-TAD}[\hat{X}_{0|0},P_{0|0},\hat{X}_{k|k-1},P_{k|k-1}]$
	\begin{itemize}
		\item Set initial values of $X_{0|0}$ and $P_{0|0}$.
		\item \textbf{for} $k=1:k_{max}$
		\begin{itemize}
			\item Calculate $S_{k-1|k-1}=chol(P_{k-1|k-1})$.
			\item Calculate $\hat{X}_{k|k-1}$ and $P_{k|k-1}$ using \eqref{eqxpredict} and \eqref{eqPpredict}.
			\item Calculate $S_{k|k-1}$ and $S_{R,k}$ using \eqref{eqchol}.
	        \item Calculate $(x_{s,k},y_{s,k})$ using \eqref{eqmomentsensorarray} and \eqref{eqdepth}.
			\item Calculate $\hat{Y}_{k|k-1}$ using \eqref{eqy}.
			\item Calculate $P_{xy,k|k-1}$ using \eqref{eqpxy}.
			\item Calculate $\hat{X}_{k|k} = FPI[\hat{X}_{k|k-1}, Y_k]$.
			\item Calculate $P_{k|k}$ using  (\ref{eqpkk}).
            \item Calculate $\sigma^*_k$ using \eqref{eqnoptimalkernel}.
		\end{itemize}
		\item \textbf{end for}
	\end{itemize}  
\end{algorithm}

\section{Simulation Results}
This section presents the implementation of a  BOT problem for an engagement scenario taken from Ref. \cite{radhakrishnan2018new,leong2013gaussian}. The target states are estimated using existing Gaussian filters, such as EKF \cite{bar2004estimation}, UKF \cite{wan2000unscented} , CKF \cite{arasaratnam2008square}, GHF \cite{ito2000gaussian}, and PCKF \cite{kumar2023polynomial}, and their corresponding maximum correntropy variants. The estimation is performed utilizing bearing data collected from the proposed dynamically modelled towed sensor-array, where the bearing data is corrupted with non-Gaussian noise. These estimation results are compared with existing BOT methods that assumes a fixed aft position for the towed sensor-array \emph{w.r.t.} stern of the tow-ship \cite{kumar2016integrated,kumar2021conditioned,jahan2020fusion}. Their performances are compared in terms of various performance metrics.  

\subsection{Engagement Scenario}
The target and tow-ship is considered to be moving on the sea surface at a constant velocities \cite{leong2013gaussian}, with the target moving in straight line and the tow-ship performs a manoeuvre from $13^{th}$ min to $17^{th}$ min. The parameters used for modelling the target and tow-ship motion is listed in Table \ref{tablescenario}.
Bearing data is collected from the towed sensor-array at sampling intervals of $10 \sec$, $12 \sec$, and $24 \sec$. The engagement scenario for target and own-ship is shown in Fig. \ref{figengage}(a).
The measurement noise in the bearing data follows a non-Gaussian probability distribution, given by $r_k\sim0.9\mathcal{N}(0,(2.5^o)^2)+0.1\mathcal{N}(0,(2.5^o\times 10)^2)$ \cite{urooj20222d}.
The physical specifications of TCSAS used in this study are provided in Table \ref{tablephysical}.
\begin{table}
	\caption{Tracking scenario parameters.}
	\begin{center} 
		\begin{tabular}{ p{4cm} p{3cm}} 
				\hline
				Parameters &  values \\
				\hline
				Target initial range & 5 $km$\\
				Target speed & 4 knots \\
				Target course & $-140^o$ \\
				Tow ship starting point & $(0,0)$ km \\
				Tow ship speed & 5 knots\\
				Tow ship initial course & $140^o$ \\
				Tow ship final course & $20^o$ \\
				Tow ship manoeuvre & $13^{th}$ to $17^{th}$ min\\
				Operation time & 30 min\\
				Process noise intensity ($\bar{q}$) & $1.944\times 10^{-6}  km^2/min^3$\\
				\hline
		\end{tabular}
		\label{tablescenario}
	\end{center}
\end{table}
\begin{table}
\caption{Physical parameters of towed cable sensor-array system \cite{guo2021numerical}.}
	\begin{center} 
		\begin{tabular}{ p{3cm} p{2cm}p{2cm} } 
			\hline
			\textbf{Parameters} & \textbf{Tow cable} & \textbf{Sensor-array}\\
			\hline
			Length $(m)$ & 723 & 273.9\\
			Diameter $(m)$ & 0.041 & 0.079\\
			Mass/length $(kg/m)$ & 2.33 & 5.07 \\
			$C_{D,n}$ & 2 & 1.8\\
			$C_{D,t}$ & 0.015 & 0.009\\
			\hline
		\end{tabular}
		\label{tablephysical}
	\end{center}
\end{table}

\subsection{Dynamics of TCSAS}
\begin{figure*}
	\centering
	\begin{subfigure}[b]{0.48\textwidth} 
		\centering
		\includegraphics[width=\textwidth]{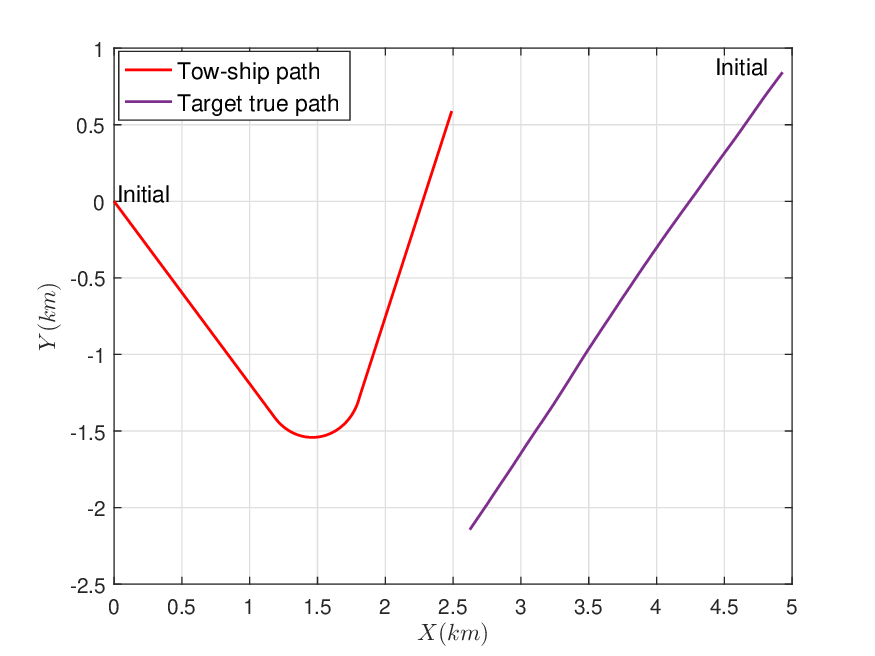}
	\end{subfigure}
	\begin{subfigure}[b]{0.48\textwidth}
		\centering
		\includegraphics[width=\textwidth]{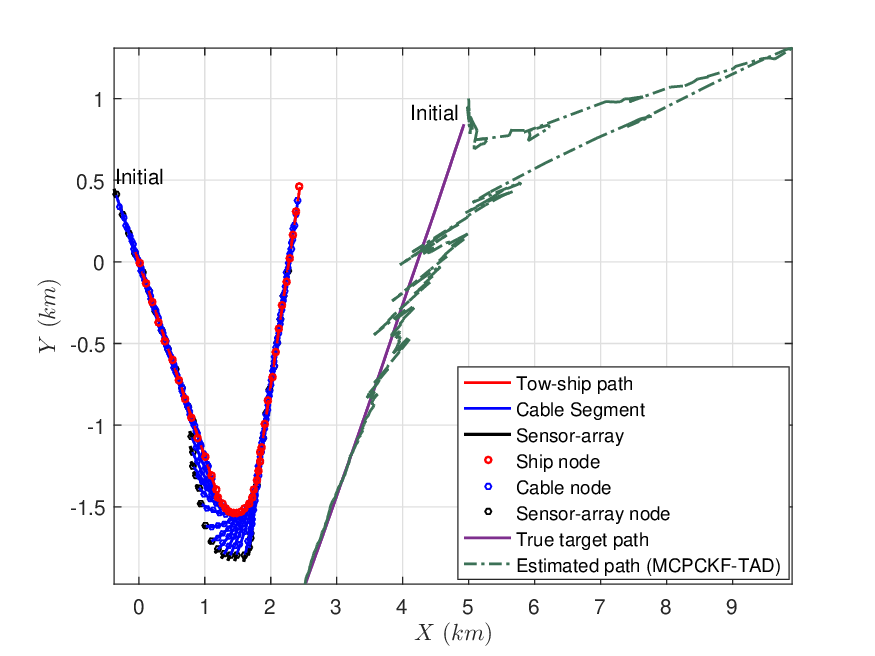}
	\end{subfigure}
	\caption{(a) Engagement scenario showcasing the path traced by tow-ship and target, and (b) estimated target path using MCPCKF-TAD.}
	\label{figengage}
\end{figure*}

The tow cable is modelled flexible using six interconnected rigid segments, with the top segment connected to tow-ship using a hinged joint, and the sensor-array is connected to the end of $6^{th}$ segment of tow cable using a revolute joint.
The dynamic analysis of TCSAS for the tow-ship motion from Fig. \ref{figengage} is performed using lumped mass approach proposed in Section \ref{sectiondynamicmodel} in Matlab 2018 software.
 For each time instant, the azimuth ($\alpha^i$) for interconnected cable segment and sensor-array is obtained by solving the ODE from Eqn. \eqref{eqmomentfinal} and Eqn. \eqref{eqmomentsensorarray}, respectively using $4^{th}$ order Runge-Kutta method \cite{press2007numerical}.
The azimuth $\alpha^i$ for each segments are passed through kinematic Eqn. \eqref{eqkinematics}, to deduce the position, velocity, and acceleration of $(i+1)^{th}$ segment in $xy$ plane, which are utilized for solving the ODE of next rigid segment.
The ODE solution demands the computation of intractable integrals which are involved in calculation of drag force from Eqn. \eqref{eqdrag}, and moment due to drag force in Eqn. \eqref{eqmomentdragintegral}.
In this study, these integrations are approximated using Gauss-Legendre five point rule, such that $\int_{0}^{L} \mathcal{F}({\mathcal{X}}) d\mathcal{X} = \textstyle\sum_{i=1}^{n} W_i \mathcal{F}(\zeta_i)$, where the $\zeta_i$ and $W_i$ are points and associated weights, respectively \cite{press2007numerical}. 
The ODE solution at t=1, demands the initial azimuth, and azimuth rate for each segments, for that the TCSAS is assumed to be initially in aft with the tow-ship, such that $\alpha_1^i=\alpha_{towship}^1$ and $\dot{\alpha}_1^i=0$ for $i=1\cdots 7$.
The boundary condition parameters such as position, velocity, and acceleration at the top-end of cable segment is same as that of tow-ship, which is known during entire operation period.
The depth ($h^i$) and elevation angle ($\beta^i$) of rigid segments is obtained by solving the force balance Eqns. \eqref{eqforcebalance}-\eqref{eqdepth}, which are deduced under the quasi-static equilibrium condition at the nodes. 
The ODE deduced from the moment balance condition, and force balance equation deduced from the quasi-static equilibrium condition, are solved simultaneously for each time instant using the procedure described above, to obtain the azimuth, elevation, and position in 3D space of interconnected rigid segments.

\begin{figure}
	\centerline{\includegraphics[scale =0.5]{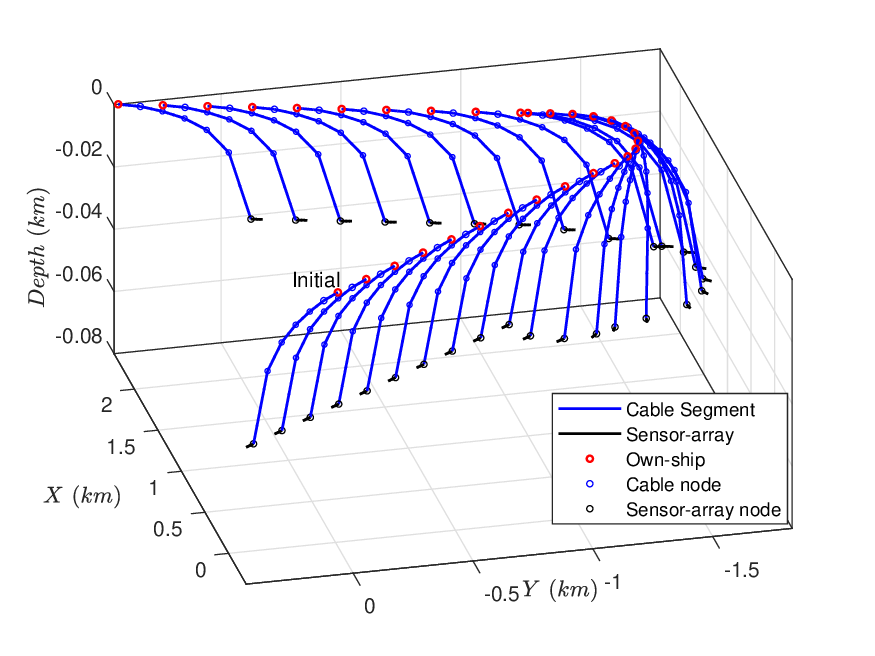}}
	\caption{Profile view of TCSAS dynamics.}
	\label{figvscenario_front}
\end{figure}

\begin{figure}
	\centerline{\includegraphics[scale =0.5]{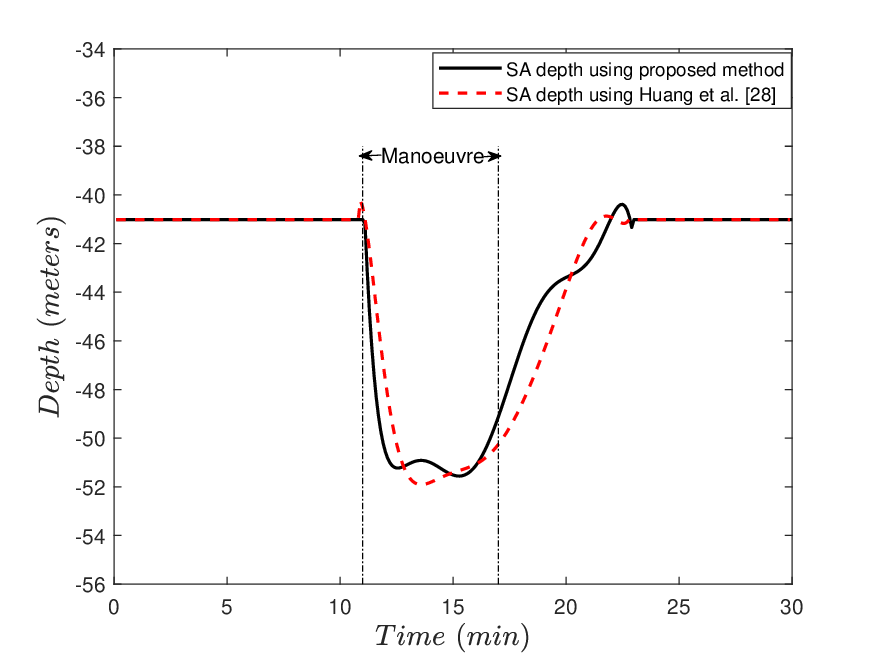}}
	\caption{Depth attained by CG of the sensor-array.}
	\label{figdepth}
\end{figure}

\begin{figure*}
	\centering
	\begin{subfigure}[b]{0.48\textwidth} 
		\centering
		\includegraphics[width=\textwidth]{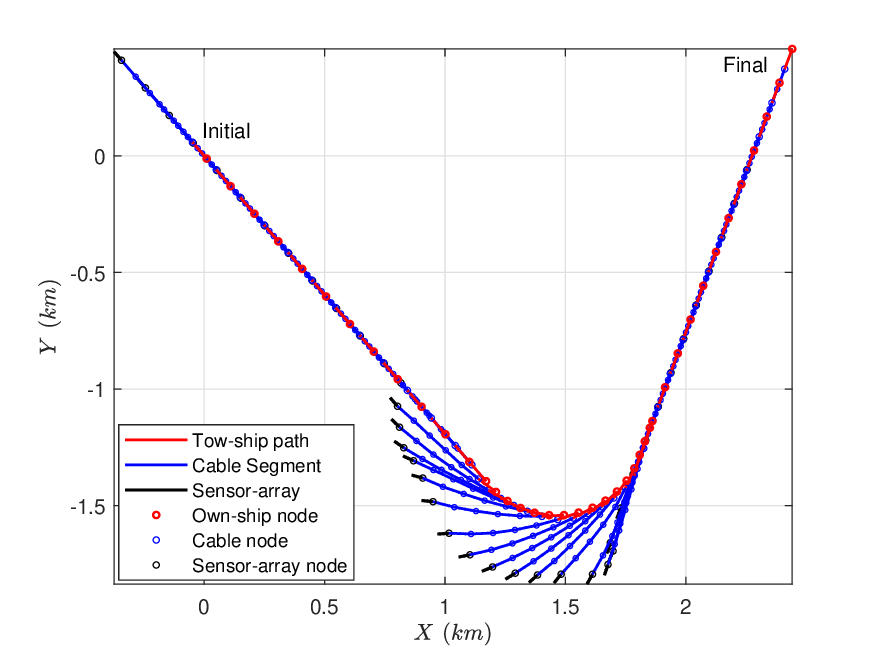}
	\end{subfigure}
	\begin{subfigure}[b]{0.48\textwidth}
		\centering
		\includegraphics[width=\textwidth]{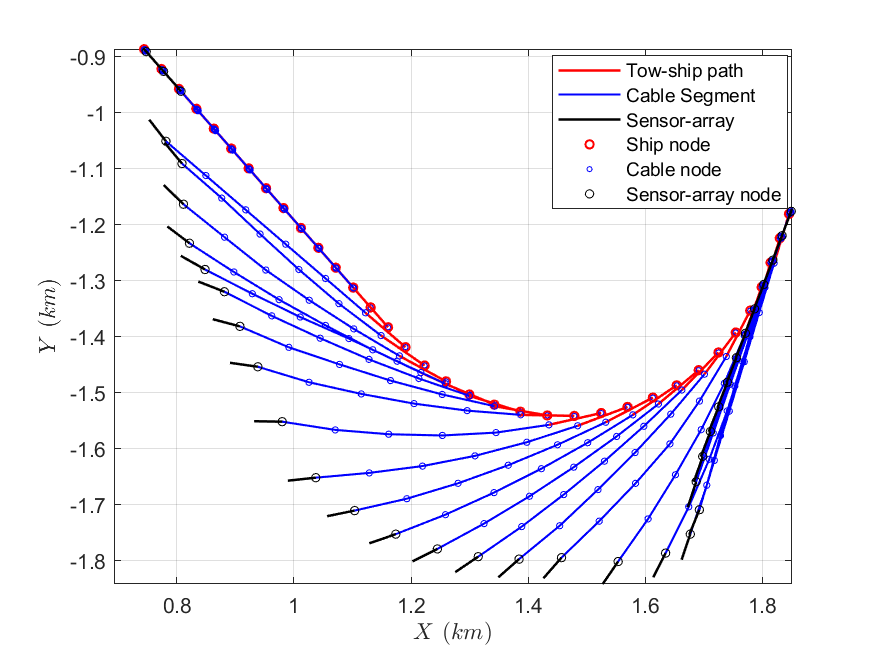}
	\end{subfigure}
	\caption{Top view of TCSAS dynamics and its enlarged view during manoeuvre for own-ship trajectory.}
	\label{figvscenario_top}
\end{figure*}

The profile view shown in Fig. \ref{figvscenario_front}, effectively illustrates the 3D dynamics of towed cable segments and sensor-array for tow-ship trajectory from Fig. \ref{figengage}(a). 
It can be observed that the depth of the towed body increases during the maneuvers due to a reduction in the  tow-ship’s forward velocity caused by its radial speed component. This reduction in forward velocity decreases the lift force acting on the towed body as verified by Eqn. \eqref{eqforcebalance}, which causes it to dive deeper until a new stable depth is achieved, which is maintained throughout the maneuvers \cite{sun2011dynamic,huang1994dynamic}.  The depth variation of the sensor-array’s center of gravity (CG) is shown in Fig. \ref{figdepth}. It shows that the sensor-array depth increases during the ship's maneuvers, then stabilizes at approximately $50m$ during the turn, and then gradually decreases as the maneuvers concludes.
The depth of the sensor-array is negligible compared to the range of target from tow-ship. Such that the bearing data can be reasonably approximated as if it were collected by a sensor-array floating on the sea surface. This approximation reduces the complexity of the BOT problem by transforming it from a 3D to a 2D framework.
The dynamics of TCSAS in $xy$ inertial reference frame (top-view) and its enlarged view during the manoeuvre is shown in Fig. \ref{figvscenario_top}.
The red, blue and black colour plots show the path traced by the tow-ship, six cable segments and a sensor-array, respectively, at each time instant. 
It can be observed that, the TCSAS remains in aft with the tow-ship while the tow-ship is moving in straight path. However, when the tow-ship initiates a manoeuvre, the towed body deviates from its original path. Following the completion of the manoeuvre, the TCSAS gradually stabilizes due to its inertia and attains a steady-state position after some delay. 
The coordinates of the sensor-array CG in $xy$ inertial reference frame are fed to the measurement update equation in Algorithm \ref{algoBOT} to perform target tracking during BOT in the 2D plane.

\begin{figure}
	\centerline{\includegraphics[scale =0.5]{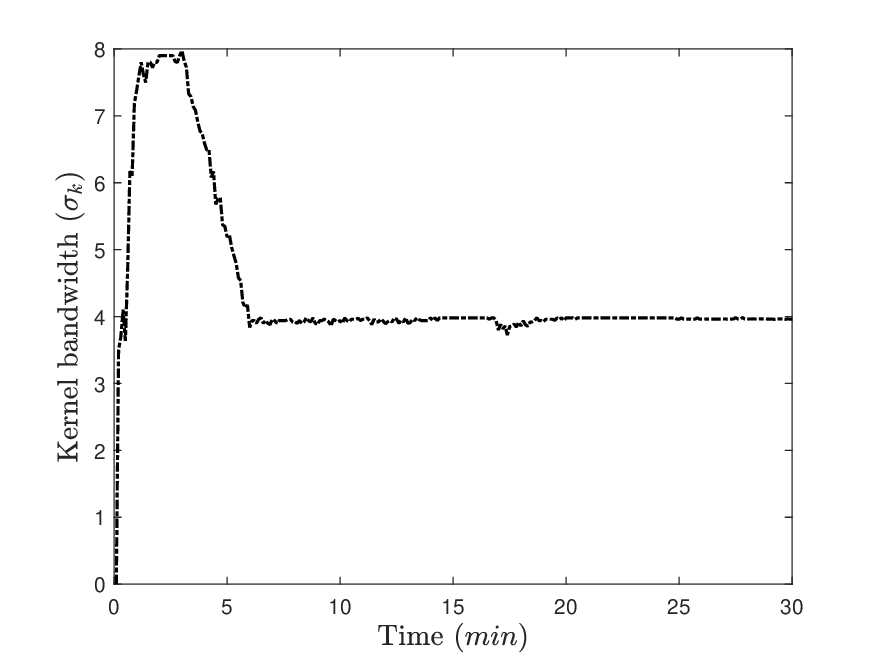}}
	\caption{Kernel bandwidth ($\sigma_k$) values with time.}
	\label{figkernel}
\end{figure}
\subsection{Performance comparison}
The BOT of a target is performed for the engagement scenario shown in Fig. \ref{figengage}(a) \cite{radhakrishnan2018gaussian,leong2013gaussian}, using the bearing data incoming from the target, which is corrupted with noise having non-Gaussian pdf. This bearing measurement is collected from a dynamically modelled towed sensor-array, which is floating on the sea surface as shown in Fig. \ref{figvscenario_top}.

The filter performance is evaluated in two key aspects. First, correntropy filters are compared with Gaussian filters to demonstrate the effectiveness of maximum correntropy filters in handling the non-Gaussian sensor noise while estimating the target states. Second, the study addresses a limitation in existing state-of-art BOT filtering approach, which assumes the sensor-array stays directly aft to the stern of the tow-ship, even during maneuvers. In reality, the towed body deviates significantly from its aft position during the tow-ship manoeuvre. To address this, the BOT is performed using the sensor data collected from the proposed dynamically modelled towed sensor-array from Section \ref{sectiondynamicmodel}.

The nomenclature used for identifying the filters reflects their specific configurations. Gaussian filters are referred to by their standard acronyms, such as EKF, UKF, CKF \emph{etc}, and the corresponding correntropy filters are denoted with the prefix ``MC" such as MCEKF, MCUKF, MCCKF \emph{etc}. The correntropy filters that use bearing data collected from the linear aft arrangement of the sensor-array \emph{w.r.t} stern of the tow-ship, without considering towed array dynamics are named as MCEKF, MCUKF, MCCKF, \emph{etc}. In contrast, filters that incorporate the dynamics of the towed sensor-array from Section \ref{sectiondynamicmodel} are labelled with the suffix ``TAD" appended to their names, such as MCEKF-TAD, MCUKF-TAD, MCCKF-TAD \emph{etc}. Gaussian filters implemented with TCSAS modelling also follow this naming convention, such as, EKF-TAD, UKF-TAD, CKF-TAD \emph{etc}.

The filtering performance is evaluated using four key metrics, which are root mean square error (RMSE), percentage track loss, average RMSE, and execution time.
The RMSE for a state $x_k$ at time $k$ calculated over $MC$ Monte Carlo runs is $({ \dfrac{1}{MC}\sum_{j=1}^{MC}{(x_{j,k}-\hat{x}_{j,k|k})^2} })^{1/2}$. 
The $\%$ track loss quantifies the reliability of a filter by checking whether the estimated target position diverges significantly from the true position at the final time step. A track loss is declared when the distance between the estimated and true positions at final time step $(k_{max})$, \emph{i.e.} $\sqrt{(x_{kmax}-\hat{x}_{kmax|kmax})^2+(y_{kmax}-\hat{y}_{kmax|kmax})^2}$, exceeds a predefined threshold called the track bound, which is taken as $250m$ in this study. Execution time measures the computational efficiency of the filters, which is determined by executing all filters under identical conditions on the same system. Filters with higher computational complexity takes longer to execute, directly reflecting their computational demand. The estimated target path using the proposed MCPCKF-TAD is shown in Fig. \ref{figengage}(b). 

\begin{figure*}
	\centering
	\begin{subfigure}[b]{0.48\textwidth} 
		\centering
		\includegraphics[width=\textwidth]{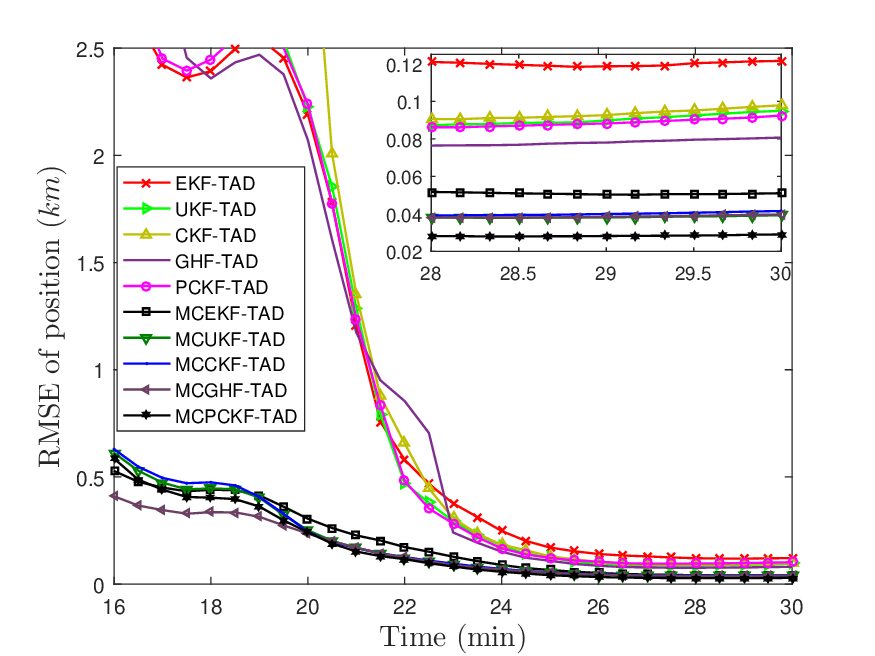}
	\end{subfigure}
	\begin{subfigure}[b]{0.48\textwidth}
		\centering
		\includegraphics[width=\textwidth]{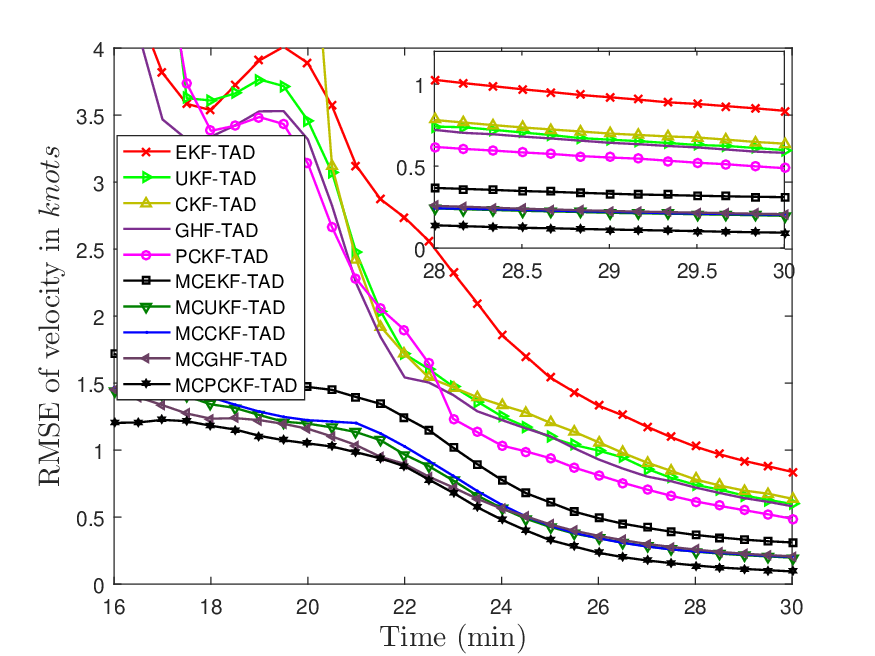}
	\end{subfigure}
	\caption{RMSE of (a) position in $km$ and (b) velocity in $knots$ ($kt$) for tracking the target using Gaussian filters and maximum correntropy filters considering towed array dynamics and non-Gaussian measurement noise.}
   \label{figrmse_Gaussian_MC}
\end{figure*}

Firstly, the performance of Gaussian filters and maximum correntropy filters are studied for the bearing data corrupted with non-Gaussian noise collected from the dynamically modelled towed sensor-array. 
The sensor-array position is extracted from Fig. \ref{figvscenario_top} and this is obtained using the towed array dynamics modelling method proposed in Section \ref{sectiondynamicmodel}. 
All the filters are initialized identically using the technique from Ref. \cite{radhakrishnan2018new,leong2013gaussian}. Maximum correntropy filters are implemented using the proposed Algorithm \ref{algoBOT} for BOT. The kernel bandwidth selected at each time step is given in Fig. \ref{figkernel}.
The RMSE plot of position and velocity estimate for 1000 Monte Carlo simulation run is shown in Fig. \ref{figrmse_Gaussian_MC}. It shows that the filters begin converging only after the tow-ship completes its maneuvers, around the $17^{th}\ min$.
It can be inferred from the RMSE plots that the Gaussian filters, based on the minimum mean square error criterion, are sensitive to non-Gaussian noise, causing their RMSE plots to converge at higher values. In contrast, maximum correntropy filters minimize both second and higher-order error moments, enabling faster convergence, as shown in the Fig. \ref{figrmse_Gaussian_MC}. Among Gaussian filters, EKF-TAD shows the highest RMSE. For maximum correntropy filters, MCPCKF-TAD achieves the fastest convergence, while sample-point filters like MCCKF-TAD, MCUKF-TAD, and MCGHF-TAD perform comparably. Table \ref{figrmse_Gaussian_MC} summarizes the filters performance in terms of $\%$ track loss, average RMSE for position and velocity, clearly demonstrating the superior performance of maximum correntropy filters in the presence of non-Gaussian noise.

\begin{figure*}
	\centering
	\begin{subfigure}[b]{0.48\textwidth} 
		\centering
		\includegraphics[width=\textwidth]{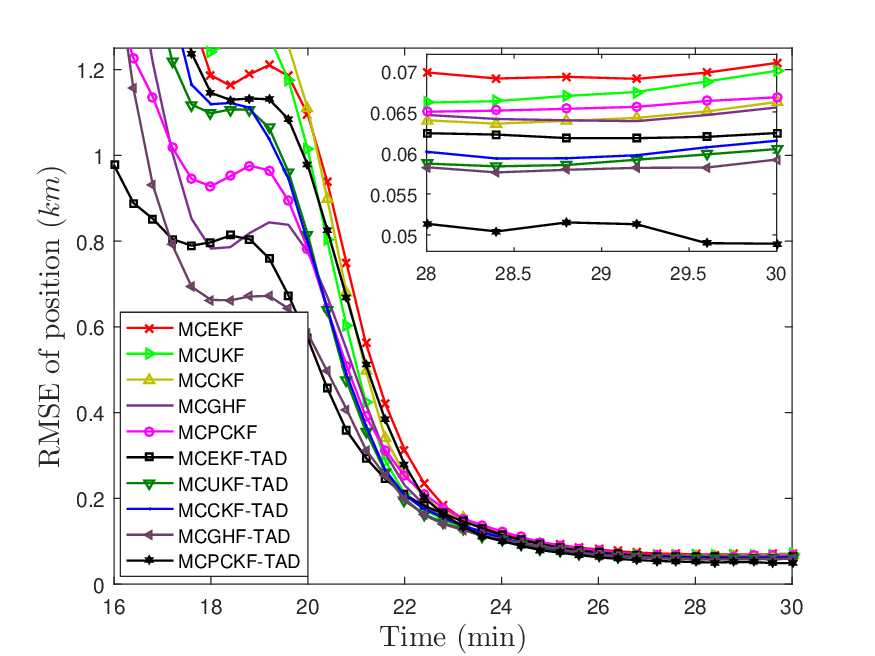}
	\end{subfigure}
	\begin{subfigure}[b]{0.48\textwidth}
		\centering
		\includegraphics[width=\textwidth]{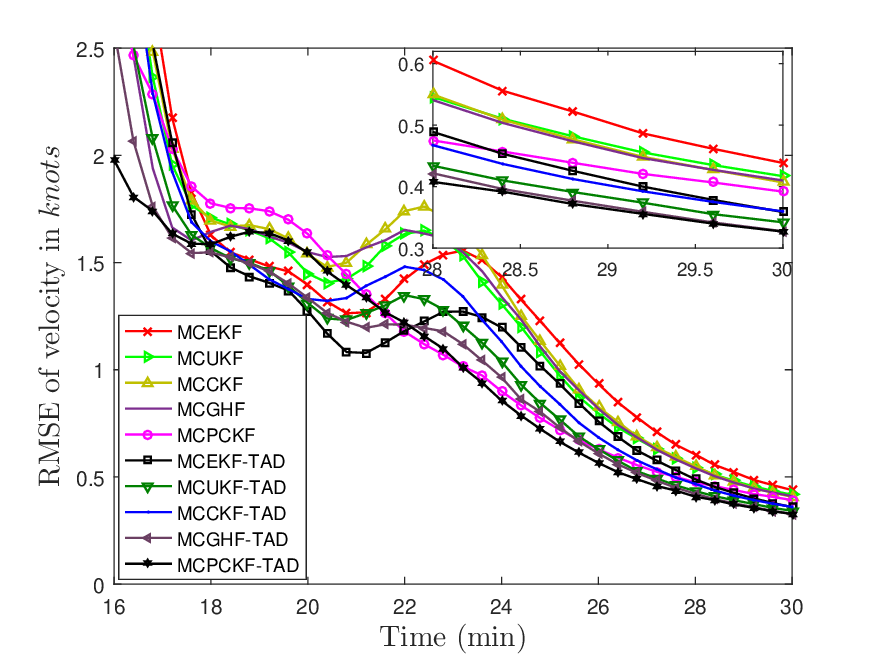}
	\end{subfigure}
	\caption{RMSE of (a) position in $km$ and (b) velocity in $knots$ ($kt$) using maximum correntropy filters with non Gaussian noise with and without dynamic model.}
	\label{figrmse_MC_linearaft_sensorarray}
\end{figure*}

From the above discussion, it can be concluded that the BOT problem with non-Gaussian sensor noise, can be handled effectively using maximum correntropy filter.
To further evaluate performance, the estimators utilizing bearing data from the proposed dynamically modelled towed sensor-array are compared with the existing BOT approach \cite{kumar2021conditioned,jahan2020fusion,kumar2016integrated,xu2017particle,kim2022surface}, where the sensor data is assumed to be collected from a fixed aft position relative to the stern of the tow-ship. The maximum correntropy based Kalman filter is applied to both approaches for performance assessment. For clarity, filters using the proposed method are denoted as MCEKF-TAD, MCCKF-TAD, etc., while those using the existing BOT methods are MCEKF, MCCKF, etc.
The RMSE of position and velocity estimate for a sampling time $T_s=12\ sec$, obtained for 1000 independent $MC$ runs excluding the diverged tracks, is shown in Fig. \ref{figrmse_MC_linearaft_sensorarray}, for both the approaches.
The results demonstrate that the proposed approach converges faster and achieves lower RMSE than existing BOT technique. Table \ref{tablearmse_MC_linearaft_sensorarray} further compares the two approaches in terms of track loss percentage, average RMSE, and execution time for different sampling intervals.
The proposed method significantly reduces track loss and achieves lower RMSE values. Among the filters with dynamically modelled TCSAS, MCPCKF-TAD performs the best and the estimated target path traced by this filter is shown in Fig. \ref{figengage}(b). However, filters with towed array dynamics require higher computations, resulting in longer execution times due to the complexity of modelling the towed body dynamics. It can be concluded from the RMSE results in Fig. \ref{figrmse_MC_linearaft_sensorarray} and average RMSE comparison Table \ref{tablearmse_MC_linearaft_sensorarray} that the dynamics modelling for TCSAS significantly improved the filtering process, leading to improved accuracy and robustness.
\begin{table*} 
	\centering 
	\caption{$\%$ Track Loss (TL), average RMSE for position (km), and velocity (knots) for Gaussian filters and maximum correntropy filters considering towed array dynamics in the presence of non-Gaussian measurement noise.}
	\begin{tabular}{|c|c|c|c|c|c|c!{\vrule width 1.1pt}c|c|c|c|c|}
		\hline 
		$T$ & Parameter & EKF-TAD & UKF-TAD & CKF-TAD & GHF-TAD & PCKF- & MCEKF- & MCUKF- & MCCKF- & MCGHF-& MCPCKF- \\
		($sec$)& & & & & & TAD&TAD & TAD&TAD &TAD & TAD\\
		\hline
		\multirow{3}{*}{10} & $\%$ TL     & 14.1 & 5.5  & 6.1 & 4.7 & 4.6    & 1.9 &  0.1  & 0.1 &  0 & 0  \\ \cline{2-12}
		& Position    & 6.93 & 5.40 & 5.61 & 5.04 & 4.84 & 2.94 & 2.19 & 2.27 & 2.00 & 1.85  \\ \cline{2-12}
		& Velocity    & 22.17 &21.09 & 21.30 & 19.82 & 18.70& 7.08&6.45 & 6.74 &  6.51& 6.28  \\ \cline{2-12}
		\hline
		\multirow{3}{*}{12} & $\%$ TL     & 15.1 & 7.8  & 7.1 & 6.6 & 6.8  & 2.2 & 0.1 & 0.3  &  0 & 0.4 \\ \cline{2-12}
		& Position    & 6.08 & 4.88 & 5.88  & 5.38 & 4.69  & 3.06 & 2.31 & 2.41 &  2.12 & 1.97  \\ \cline{2-12}
		& Velocity    & 22.03 &18.48&  21.03 & 20.39 & 18.24& 11.41& 9.76 & 10.16& 8.84 &  8.58 \\
		\hline
		\multirow{3}{*}{24} & $\%$ TL     & 22.3 & 14.1 & 13.7 & 11.5 & 10.3  & 2.9 & 1.7 & 1.3 & 1.7 & 0.9  \\ \cline{2-12}
		& Position    & 4.87 & 4.19 & 4.10 & 4.22 & 3.46  & 3.21 & 2.85 & 2.92 &   2.64 & 2.30  \\ \cline{2-12}
		& Velocity    & 15.92 & 15.37 & 14.87 & 15.28 & 12.51 & 12.93& 11.29& 11.42& 9.88 &9.071  \\ \cline{2-12}
		\hline
	\end{tabular}
	\label{tablearmse_Gaussian_MC}
\end{table*}
\begin{table*} 
	\centering 
	\caption{$\%$ Track Loss (TL), average RMSE of position (km) and velocity (knots) for maximum correntropy-based filters using bearing data with a non-Gaussian noise statistics, collected from a sensor-array with and without dynamic model.}
	\begin{tabular}{|c|c|c|c|c|c|c!{\vrule width 1.1pt}c|c|c|c|c|}
		\hline 
		$T$ & Parameter & MCEKF & MCUKF & MCCKF & MCGHF & MCPCKF & MCEKF- & MCUKF- & MCCKF- & MCGHF-& MCPCKF- \\
		($sec$)& & & & & & &TAD & TAD&TAD &TAD & TAD\\
		\hline
		\multirow{3}{*}{10} & $\%$ TL   & 1.6 & 0.1 &  0.2 &  0.1 & 0 & 1.9 & 0.1 &  0.1 &  0 &  0  \\ \cline{2-12}
		& Position    & 3.14 & 3.00 & 2.97 & 2.85 & 2.80 &  2.54 &  2.18 & 2.27 &  2.00 & 1.85  \\ \cline{2-12}
		& Velocity    &12.29 & 10.60 & 10.98 & 9.77 & 9.53 & 10.08 & 9.45 & 9.74 & 8.51  & 8.28  \\
		\hline
		\multirow{3}{*}{12} & $\%$ TL     & 1.4 & 0.2 & 0.2 & 0.1 & 0.1& 2.2  & 0.1& 0.3  & 0.4 & 0 \\ \cline{2-12}
		& Position    & 3.48 & 3.10&  3.04 & 2.97 & 2.89 &   2.66 &  2.31 &  2.40 &  2.12 & 1.96  \\ \cline{2-12}
		& Velocity    &12.97 & 11.14 & 11.27 & 10.10 & 9.78 &  11.41 & 9.76 & 10.16 & 8.84 &  8.58    \\
		\hline
		\multirow{3}{*}{24} & $\%$ TL  & 2.7 & 1.9 & 1.6 & 2.2 & 1.6&   2.9 &1.7 & 1.3 & 1.7 &  0.9 \\ \cline{2-12}
		& Position    & 3.83 & 3.27  & 3.24  & 3.15  & 3.10  &    3.05   &  2.85   &  2.92  &  2.64 &  2.29    \\ \cline{2-12}
		& Velocity    & 15.91 & 12.57 & 12.43 & 10.92 & 10.23 &  10.93 & 9.29 & 9.42 & 7.88 &  7.07  \\
		\hline
		\multicolumn{2}{|c|}{Exec. time ($sec$)}& 6 &11.16 & 11.29 & 20.85 & 11.44& 197.8 & 203& 204.1& 212.6 & 203.87\\ \hline
	\end{tabular}
	\label{tablearmse_MC_linearaft_sensorarray}
\end{table*}

\section{Conclusion}
This work addresses two key challenges encountered during passive bearing-only tracking: (i) accurately determining the position of the towed sensor array during tow-ship manoeuvres, and (ii) addressing the effect of non-Gaussian noise in bearing data within the Kalman filtering algorithm. Traditional BOT methods assume a fixed aft position for the towed sensor-array, which leads to unreliable state estimation during the tow-ship manoeuvres. To overcome this limitation, a physics-based dynamic model of the TCSAS using Newtonian mechanics is proposed in this paper. The cable is modelled as flexible by discretising it into interconnected segments. The configuration of the system is obtained by solving equations derived from moment balance and quasi-static equilibrium conditions at each node. This provides a complete 3D representation of the TCSAS, providing accurate location of the sensor-array, which is further fed to the state estimation algorithm for target tracking.
Additionally, a maximum correntropy-based Kalman filter with a novel kernel bandwidth selection technique is presented to address the effect of non-Gaussian measurement noise.
The proposed method is validated for a real-world BOT engagement scenario. Results show that incorporating dynamically modelled sensor-array data into the estimation algorithm significantly improves tracking accuracy. Furthermore, it is found that the MCC-based filter outperforms conventional MMSE-based filters in handling non-Gaussian noise.


\bibliographystyle{IEEEtran}
\bibliography{refer}
\end{document}